\documentclass{article}

\usepackage{arxiv_ms/arxiv}

\usepackage[utf8]{inputenc} % allow utf-8 input
\usepackage[T1]{fontenc}    % use 8-bit T1 fonts
\usepackage{hyperref}       % hyperlinks
\usepackage{url}            % simple URL typesetting
\usepackage{booktabs}       % professional-quality tables
\usepackage{amsfonts}       % blackboard math symbols
\usepackage{nicefrac}       % compact symbols for 1/2, etc.
\usepackage{microtype}      % microtypography
\usepackage{lipsum}
\usepackage{graphicx}
\usepackage{amsmath}
\graphicspath{ {./images/} }

\usepackage{color}
\usepackage{caption}
\usepackage{orcidlink}

\title{%Elementary processes of 
Grain boundary evolution in nanoparticles}

\author{
 Manoj Settem \begingroup
\hypersetup{hidelinks}
\large{\orcidlink{0000-0001-7287-6827}}
\endgroup  \\
  Dipartimento di Ingegneria Meccanica e Aerospaziale\\
  Sapienza Università di Roma\\
  via Eudossiana 18, Roma, Italy, 00184 \\
  \texttt{manoj.settem@uniroma1.it} \\
   \And
 Pranav Kumar \begingroup
\hypersetup{hidelinks}
\large{\orcidlink{0000-0002-3661-5870}}
\endgroup  \\
  Institute for Materials Science\\
  University of Stuttgart\\
  Stuttgart, Germany, 70569 \\
  \And
 Ajeet K. Srivastav \begingroup
\hypersetup{hidelinks}
\large{\orcidlink{0000-0001-8986-6502}}
\endgroup  \\
  Department of Metallurgical and Materials Engineering\\
  Visvesvaraya National Institute of Technology\\
  Nagpur, India, 440010 \\
}

\date{}
\renewcommand{\today}{}

\begin{document}
\maketitle
\begin{abstract}

Grain boundary evolution is a key dynamical process that enables structural rearrangements in nanoparticles and drives them towards low energy configurations. Grain boundaries can also enhance catalytic properties, making it important to understand the elementary processes underlying their evolution for grain boundary engineering in nanoparticles. Compared with bulk materials, nanoparticles have additional rotational and translational degrees of freedom, can accommodate structural changes through shape relaxation, and grain boundaries terminate at free surfaces. However, the atomic-scale mechanisms of GB evolution under these less constrained conditions remain comparatively less understood than in bulk materials. Here, using atomistic simulations, we focus on low energy $\Sigma3$ (coherent twin boundary) and $\Sigma11$ grain boundaries, which are among the persistent compact GB structures that emerge during nanoparticle structural evolution. We identify two fundamental atomic displacements, column shift (C) and screw shift (S), that recur during the evolution of these grain boundaries and their junctions. These displacements occur through different atomic pathways and combine in different ways to generate grain boundary migration, structural transformations, and junction evolution. In particular, the same initial and final grain boundary configurations can be connected through different atomic pathways, and a column shift can occur either as a full shift or through disconnection kinks. C and S remain identifiable even when the grain boundary character changes, for example during a $\Sigma11$ to $\Sigma3$ transformation. An understanding of these elementary GB processes can help identify strategies to control grain boundary evolution and thereby engineer GB structures in nanoparticles.

%These results provide an atomic displacement based description of grain boundary evolution in nanoparticles and suggest ways to influence the active processes through nanoparticle geometry, surfaces, and local chemistry.

\end{abstract}

% keywords can be removed

\section{Introduction}
\label{sec:intro}
Metal nanoparticles can adopt single-crystalline fcc (for example, truncated-octahedral), decahedral (Dh), icosahedral (Ih), and twinned morphologies \cite{baletto2005Rev}. Thermodynamic competition among them dictates the equilibrium structural distribution (occurrence fraction of the various morphologies), which depends on temperature, particle size, metal system, and substrate \cite{settem2022Au,settem2023AgCuAu,settem2024_Au_substrate}. Coexistence of these morphologies has also been observed experimentally \cite{wells2015AuImagingACFraction,foster2018AuImagingACFraction}. Dh, Ih and twinned particles contain coherent twin boundaries (CTBs), or $\Sigma3$\{111\} grain boundaries (GBs). CTBs are therefore not merely defects to be eliminated, but intrinsic structural features characteristic of low-energy nanoparticle configurations.

Nanoparticle synthesis often involves attachment, coalescence, and growth, during which GB migration plays a key role in enabling the particle to reach lower-energy stable or metastable configurations \cite{ingham2011_Au_coalescence,lange2020,gu2024_NP_disconnection}. On the other hand, GBs can be desirable since they provide under-coordinated atomic environments, introduce strain, and modify local electronic structure, which contribute to enhanced catalytic activity in nanoparticles \cite{feng2015_gb_catalysis,mariano2017_gb_catalysis,kim2016_gb_catalysis}. However, retaining GBs remains a challenge as they have a tendency to migrate out of the particle. One strategy is to collide randomly oriented nanoparticles to form GB-rich nanoassemblies \cite{geng2024_nano_assemblies}. Another potential strategy is to explore whether GB migration can be controlled by chemically altering the nanoparticles, for instance, through alloying. A key step towards GB engineering in nanoparticles is therefore to understand the fundamental atomic-scale mechanisms underlying GB migration.

Grain boundary migration (GBM) is commonly described using disconnections, which are line defects carrying both a Burgers vector and a step height \cite{han2018_disconnection}. Disconnection-mediated GBM has been imaged experimentally across multiple scales, including poly-crystalline materials, nano-bicrystals, and nanoparticles \cite{gu2024_NP_disconnection,tian2024_Pt_disconnection,zhu2019_Au_disconnection}. Atomistic studies have further revealed GBM mechanisms including kinks in disconnection lines \cite{combe2016_disconnection_kinks} and cooperative structural rearrangements \cite{rajabzadeh2013_GB_shuffle,wei2021_atomistic_GBM}. However, these studies are largely focused on bulk materials or specific GBs, and GBM in nanoparticles remains comparatively less understood. Unlike bulk poly-crystalline materials, nanoparticles are less constrained in the sense that they are not surrounded by other grains, allowing significant grain rotation and shape relaxation, while GBs also terminate at free surfaces. These additional degrees of freedom can influence the atomic-scale processes through which GB migration and transformation occur. 

This structural freedom can also favor GB configurations consisting of compact structural units \cite{settem2024_GBJ_disclinations}. For instance, a $\Sigma3-\Sigma3-\Sigma11$ grain boundary junction (GBJ) disclination, despite the associated volumetric strain, has lower excess energy than the expected $\Sigma3-\Sigma3-\Sigma9$ GBJ. The $\Sigma3$ and $\Sigma11$ GBs are composed primarily of compact tetrahedra and capped trigonal prism (CTP) polyhedral structural units, respectively, with low excess volume \cite{banadaki2017_polyhedral_GB}. Our coalescence simulations of randomly oriented Pt and Ag nanoparticles show that, following the initial attachment and transient rearrangement, the persistent GB structures are primarily $\Sigma3$, $\Sigma11$, and their junctions (Supplementary Table \ref{sup_tab:disclination_fraction2} and Supplementary Figure \ref{sup_fgr:rand_coal_structs}), although their relative proportions depend strongly on the metal system. This preference for low energy grain boundaries is consistent with experimental studies of Au nanoparticles and nanoparticle assemblies, where $\Sigma3$ misorientations are strongly represented and $\Sigma11$ is also observed \cite{geng2024_nano_assemblies,zhu2022_GBCD_AuNP}. We therefore focus on the elementary processes governing the evolution of these compact GBs and their junctions. We identify a small set of fundamental atomic displacements, which occur and combine in different ways to give rise to the recurring GB processes observed here. In contrast to migration through multiple metastable GB configurations reported previously \cite{wei2021_atomistic_GBM}, here we show that the same initial and final compact structural configurations can be connected through multiple atomic-scale pathways.

\begin{figure}[!t]
\centering
  \includegraphics[width=1.0\textwidth]{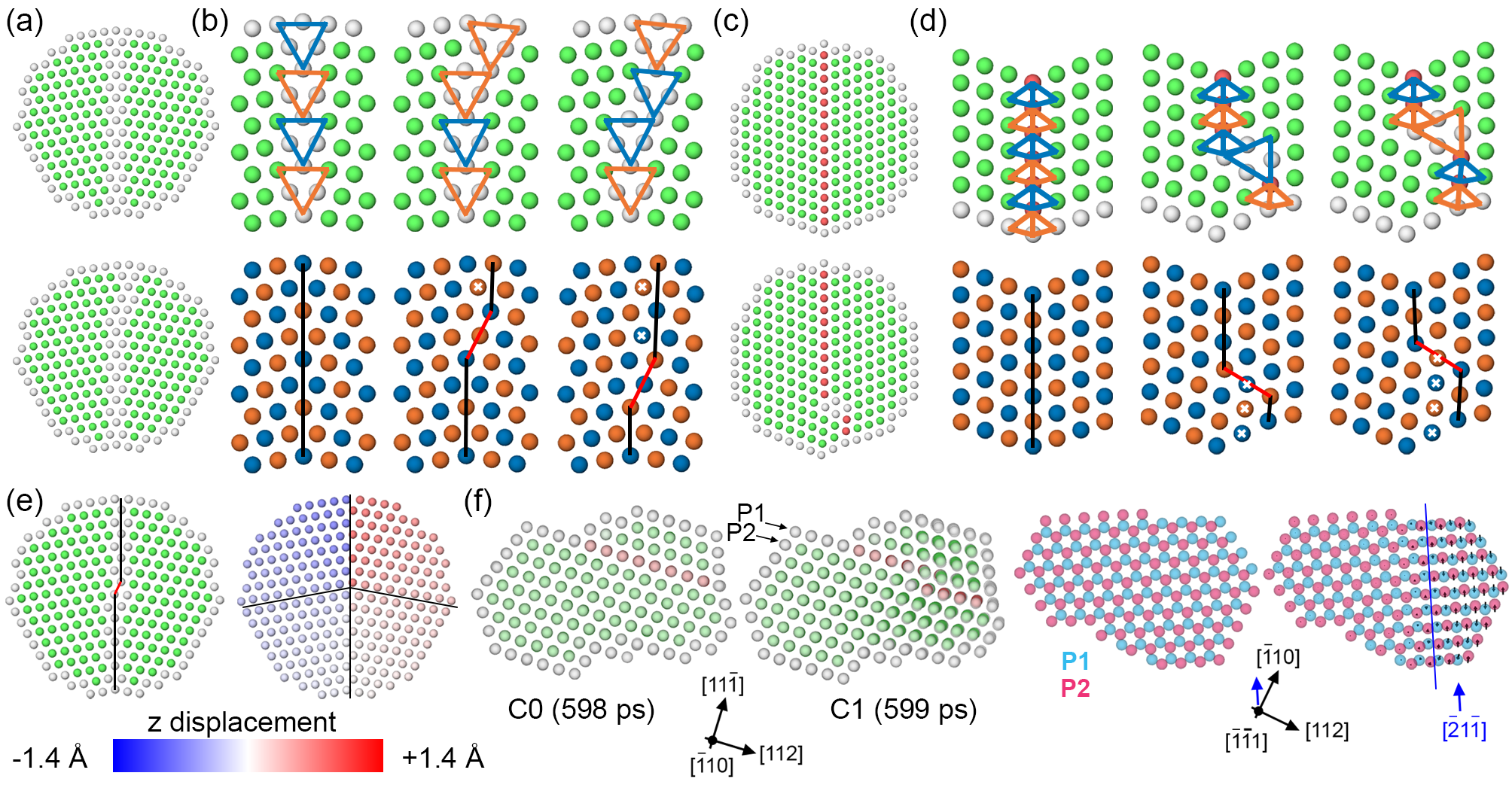}
  \caption{(a) Symmetric $\Sigma11$ grain boundary in a 2~nm diameter cylinder (top) without and (bottom) with a double disconnection. (b) Generation and migration of the double disconnection (in red) in panel (a) through two successive stacking-change column shifts (indicated by $\times$). The top and bottom images show the atomic coordination and stacking, respectively. Each CTP is represented by a downward-pointing triangle and colored according to the stacking of its three central atomic pairs. (c) Symmetric $\Sigma3$ grain boundary in a 2~nm diameter cylinder (top) without and (bottom) with a double disconnection. (d) Generation and migration of the double disconnection (in red) in panel (c) through successive stacking-change column shifts (indicated by $\times$). The top and bottom images show the atomic coordination and stacking, respectively. The disconnection-free grain boundary is composed of face-sharing tetrahedra, which are colored according to the stacking of the vertex atom not lying in the GB plane. Each CTP is represented by a left-pointing triangle. (e) Structure and $z$-displacement map of a single disconnection in the $\Sigma11$ GB generated by a screw shift. (f) Configurations C0 and C1 viewed along the tilt axis ($[\bar{1}10]$) and the grain boundary plane normal ($[\bar{1}\bar{1}1]$), showing the atomic displacements associated with the screw shift resulting in the single disconnection. In the projection along the grain boundary plane normal, only the two \{111\} planes P1 and P2 are shown, and the arrows indicate the atomic displacement vectors in C1 with respect to C0. The blue line parallel to $[\bar{2}1\bar{1}]$ indicates the disconnection line direction.}
  \label{fgr:disconnection_struct}
\end{figure}

\section{Results}
\label{sec:results}

\subsection{Fundamental atomic displacements in $\Sigma11$ and $\Sigma3$ grain boundaries}

We demonstrate the fundamental atomic displacements associated with disconnection formation and migration in $\Sigma11$ and $\Sigma3$ GBs using a cylindrical model. Figure \ref{fgr:disconnection_struct}a shows a symmetric $\Sigma11$ GB without and with a double disconnection (\emph{dd}). The $\Sigma11$ GB consists of staggered capped trigonal prisms (CTPs) \cite{banadaki2017_polyhedral_GB}, represented using different colors in Figure \ref{fgr:disconnection_struct}b. A stacking-change shift of an atomic column (indicated by $\times$) translates the topmost CTP to the right and nucleates a \emph{dd}. The column displacement is along the $\langle110\rangle$ tilt axis by one interplanar distance ($\sim1.4$ {\AA} for the Pt potential used here). A subsequent column shift migrates the disconnection downwards.When the remaining columns in the same \{113\} plane also undergo the stacking-change column shift, the GB migrates laterally by two \{113\} planes, equal to the step height of the \emph{dd}. Similar column shifts have previously been identified during shear-coupled migration of $\Sigma11$ GBs \cite{wang2026_Pt_GBM}. The same type of column shift also enables transitions to asymmetric $\Sigma11$ segments containing CTPs with the same stacking along the GB (Supplementary Figure \ref{sup_fgr:as11_displ}).

The same stacking-change column shift creates and migrates a \emph{dd} in the $\Sigma3$ GB (Figure \ref{fgr:disconnection_struct}c,d). A defect-free $\Sigma3$ GB consists of face-sharing tetrahedra, whereas at the \emph{dd} the tetrahedra are connected through a CTP. Figure \ref{fgr:disconnection_struct}d shows this configuration after three column shifts; the next column shift moves the CTP and the \emph{dd} upwards. Thus, although the defect-free $\Sigma3$ and $\Sigma11$ GBs consist of different structural units, the CTP provides a local structural connection between them during disconnection formation and migration.

A single disconnection (\emph{sd}) can instead be generated by a relative stacking-change shift of regions on either side of the GB, with one region displaced up and the other down (Figure \ref{fgr:disconnection_struct}e). In $\Sigma11$, this displacement is along the $\langle110\rangle$ tilt axis. The displacement field of the \emph{sd} is more extended than that of the \emph{dd} (Supplementary Figure \ref{sup_fgr:sd_dd_displ_maps}), and for the same line length the \emph{sd} has a higher nucleation energy (Supplementary Figure \ref{sup_fgr:discon_energetics}). A similar relative displacement produces an \emph{sd} in $\Sigma3$, but here the displacement occurs along an in-plane $\langle112\rangle$ direction, $30^\circ$ from the $\langle110\rangle$ tilt axis. Since the $\Sigma3$ \emph{sd} could not be stabilized in the cylindrical model, Figure \ref{fgr:disconnection_struct}f uses a configuration obtained from a coalescence simulation. C0 and C1 denote the defect-free and \emph{sd} configurations, shown along $\langle110\rangle$ and $\langle111\rangle$. In the $\langle111\rangle$ projection, the disconnection line is parallel to $[\bar{2}1\bar{1}]$, while the relative displacement is identified from the \{111\} planes P1 and P2. The corresponding disconnection lines, Burgers circuits, and line characteristics are provided in Supplementary Figures \ref{sup_fgr:s3_sd_dd_disconn_lines}--\ref{sup_fgr:burgers_circuit_s3} and Supplementary Table \ref{sup_tab:disconn_type}. If the stacking change is confined to a region on only one side of the GB, a single disconnection coupled to a stacking fault is produced (Supplementary Figure \ref{sup_fgr:faulted_disconn}).

These two atomic displacements form the basis of the GBM processes observed here for $\Sigma11$, $\Sigma3$, and their junctions. The first is a column shift (C), which changes the stacking of an individual atomic column. The second is a regional shift that resembles the displacement field of a screw dislocation, and we therefore refer to it as a screw shift (S). Here, ``screw shift'' refers only to the atomic displacement and does not imply that the resulting disconnection has screw character.

\begin{figure}[!t]
\centering
  \includegraphics[width=1.0\textwidth]{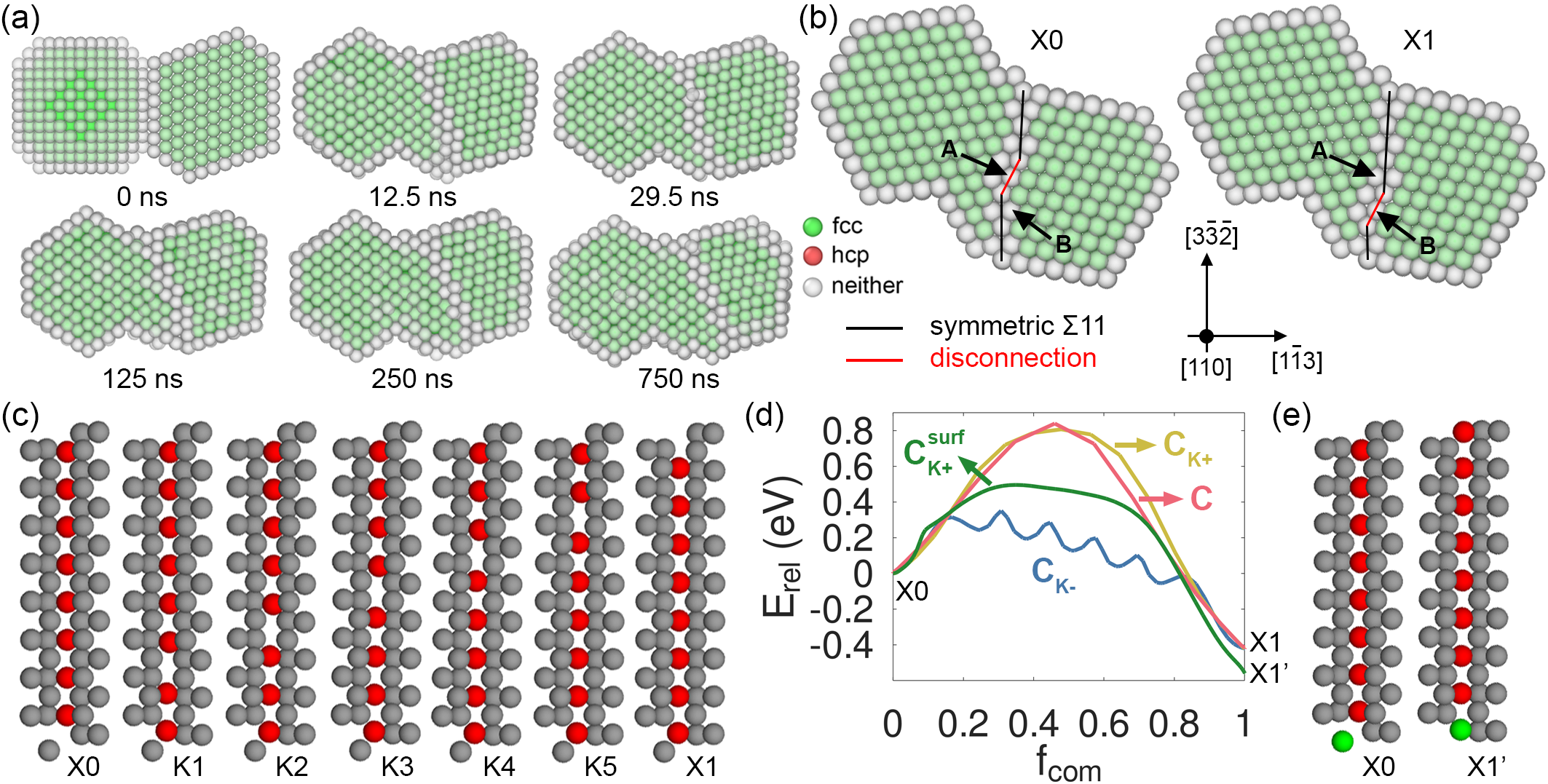}
  \caption{(a) $\Sigma11$ migration observed during coalescence of Pt$_{711}$ (particle on the left) and Pt$_{807}$ (particle on the right) nanoparticles. (b) Configurations X0 and X1 showing the downward migration of a straight-line double disconnection from column A to B. (c) Anti-parallel kink transmission viewed along $[\bar{3}32]$. The atomic column A (in red) undergoes down shift along $[\bar{1}\bar{1}0]$, and the six nearest-neighbor columns are shown in gray. (d) Energy paths for the different column-shift modes: anti-parallel kink (C\textsubscript{K-}), parallel kink (C\textsubscript{K+}), full shift (C), and parallel kink triggered by surface atom diffusion (\( \mathrm{C}_{\mathrm{K}}^{\raisebox{0.50ex}{\scriptsize surf}} \)). (e) Initial (X0) and final (X1$^{\prime}$) configurations for column shift through a parallel kink triggered by surface atom diffusion (green atom).}
  \label{fgr:disconnection_kinks}
\end{figure}

\subsection{Disconnection kinks during a column shift}
Under dynamical conditions, the atomic displacements can proceed through multiple pathways. Here, we demonstrate this for a column shift. Figure \ref{fgr:disconnection_kinks}a shows the coalescence of Pt$_{711}$ and Pt$_{807}$ nanoparticles attached at \{100\} and \{111\} facets at 850 K. A $\Sigma11$ GB forms and undergoes back-and-forth migration before settling into a symmetric $\Sigma11$ arrangement. To closely study an isolated column shift, we carried out short 10 ps simulations starting from an intermediate configuration at 29.5 ns, with configurations sampled every 10 fs. In some of these simulations, we observed migration of a straight line \emph{dd}. The relaxed energy profile (quenched to 0 K) from one such simulation is shown in Supplementary Figure \ref{sup_fgr:energy_1518_r2}. Configurations X0 and X1 represent two neighboring minima of the \emph{dd} (Figure \ref{fgr:disconnection_kinks}b), where the disconnection line is nominally located at atomic columns A and B, respectively.

The transition from X0 to X1 occurs through a stacking-change down shift of atomic column A (Figure \ref{fgr:disconnection_kinks}c). However, the entire column does not shift simultaneously. Instead, individual atoms in the column can shift separately, except for atomic pairs at the two ends of the column which shift together. In K1, shifting of the two bottom atoms creates a stacking mismatch vacancy, where two consecutive sites along the column are empty. This is different from a missing-atom vacancy, for which three consecutive sites would be empty. The stacking mismatch represents a kink, below which the disconnection line has advanced from A to B. As other atoms in the column undergo the stacking-change shift, the kink moves along the column. This motion need not be monotonic. For example, a trajectory can move from K1 to K2 to K3, return to K2, and then proceed again through K3, K4 and K5 before reaching X1. Overall, the kink migrates opposite to the column-shift direction, and we refer to this as an anti-parallel kink. The dynamical sequence of the atomic shifts is quantified from the crossover in mean distances to reference \{110\} layers on either side of the shifting column (Supplementary Figure \ref{sup_fgr:displ_mag_L1_L4}b).

%The column shift can occur in three modes: anti-parallel kink (C\textsubscript{K-}), parallel kink (C\textsubscript{K+}), and full shift (C), in which the entire column shifts simultaneously. Their atomic evolution is shown in Supplementary Figure \ref{sup_fgr:neb_images_1518}. In the parallel kink pathway, the shift instead begins near the top of the column and the kink migrates downwards, in the same direction as the column shift. A dynamically observed parallel kink sequence is shown in Supplementary Figure \ref{sup_fgr:displ_mag_L1_L4}c. Nudged elastic band (NEB) calculations comparing the three pathways are shown in Figure \ref{fgr:disconnection_kinks}d. The forward barriers follow the order anti-parallel kink $<$ parallel kink $<$ full shift.

The column shift can occur in three modes: anti-parallel kink (C\textsubscript{K-}), parallel kink (C\textsubscript{K+}), and full shift (C), in which the entire column shifts simultaneously. Their atomic evolution is shown in Supplementary Figure \ref{sup_fgr:neb_images_1518}. In the parallel kink pathway, the shift instead begins near the top of the column and the kink migrates downwards, in the same direction as the column shift. A dynamically observed parallel kink sequence is shown in Supplementary Figure \ref{sup_fgr:displ_mag_L1_L4}c. The energy barriers for the three pathways are compared using nudged elastic band (NEB) calculations in Figure \ref{fgr:disconnection_kinks}d. For the full shift, additional harmonic restraints between neighboring atoms in the shifting column are required to retain the concerted pathway during NEB; without these restraints, the pathway evolves into a parallel kink. The forward barriers follow the order anti-parallel kink $<$ parallel kink $<$ full shift.

In the 70 simulations carried out from the intermediate configuration, we observed one conventional parallel kink pathway, while the full-shift pathway was not observed. In two additional instances, we observed a parallel kink triggered by diffusion of a surface atom from a neighboring column into the column undergoing the shift (green atom in Figure \ref{fgr:disconnection_kinks}e). This results in a column up shift (\( \mathrm{C}_{\mathrm{K}}^{\raisebox{0.50ex}{\scriptsize surf}} \) in Figure \ref{fgr:disconnection_kinks}d), with the detailed evolution shown in Supplementary Figure \ref{sup_fgr:surf_diff_par_kink}. The surface diffusion appears as a bump in the energy path and lowers the forward barrier to 0.495 eV, compared with 0.807 eV for the conventional parallel kink and closer to the 0.351 eV barrier of the anti-parallel kink.

Thus, the same X0 to X1 column shift can occur through anti-parallel kink, parallel kink, or full shift pathways. Surface diffusion provides an additional route involving a parallel kink and leads to a different final configuration X1$^{\prime}$, although the \emph{dd} migrates by the same distance. These results show that even a single C displacement does not have a unique atomic pathway.

\subsection{Recurring atomic processes during grain boundary evolution}

\begin{figure}[!b]
\centering
  \includegraphics[width=1.0\textwidth]{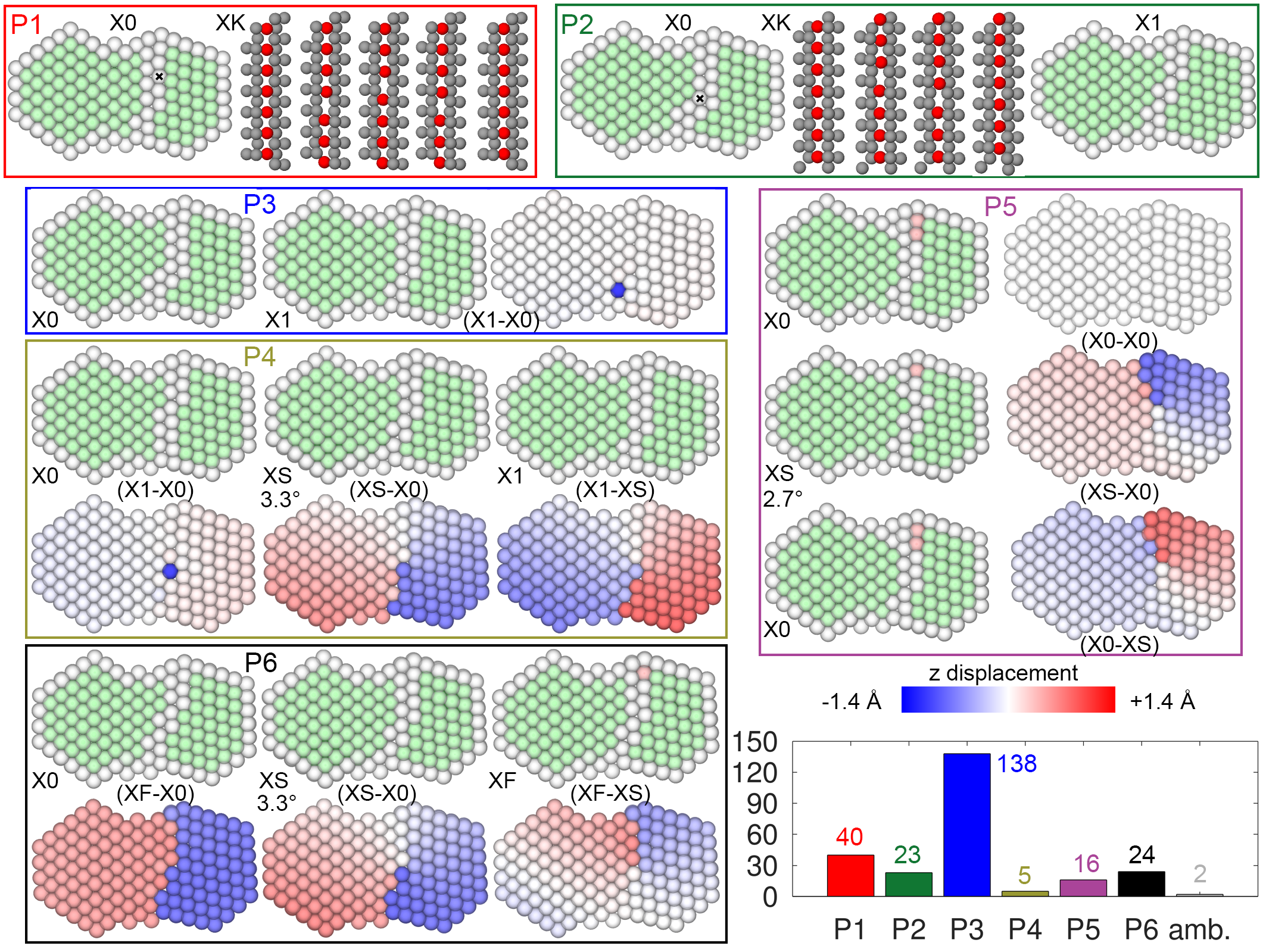}
  \caption{The six GB processes (P1 to P6) identified from 100 simulations initiated from configuration X0 shown in P2 and P3. The $z$-displacement map of configuration B with respect to A is denoted as (B-A). For processes involving XS type configurations, the corresponding twist angle is also shown. The bar plot gives the counts of each GB process in the first 50 simulations.}
  \label{fgr:gb_processes}
\end{figure}

During GB evolution, we observe a few recurring processes that can be described using combinations of the fundamental atomic displacements discussed above. To demonstrate this, we consider an intermediate configuration during the coalescence of two Pt$_{405}$ nanoparticles. The configuration contains a faceted $\Sigma11$ GB consisting of symmetric and asymmetric segments (configuration X0 in processes P2 and P3 in Figure \ref{fgr:gb_processes}). We carried out 100 independent 5 ps simulations starting from this configuration at 850 K, with configurations sampled every 10 fs to capture the different GB processes. We identified six recurring processes, labeled P1--P6, and describe them using the following notation. X0 denotes the initial configuration of a given process, XK an anti-parallel kinked configuration, X1 a configuration with one column shift relative to X0, XS a configuration with a partial screw shift, and XF a configuration with a full screw shift, where one region has shifted relative to the other by the complete stacking-change displacement. We emphasize that X0 does not represent one unique atomic configuration, but simply the initial configuration for the process under discussion. The same applies to XK, X1, XS, and XF.

The six processes can be broadly separated into column shift related (P1--P3) and screw shift related (P4--P6) processes. In P1 (X0 $\rightarrow$ XK $\rightarrow$ X0), a column shift is initiated through an anti-parallel kinked configuration but reverses before the shift is completed. When the column shift is completed through the anti-parallel kink, we denote the process as P2 (X0 $\rightarrow$ XK $\rightarrow$ X1). P3 is a full column shift without a kink (X0 $\rightarrow$ X1). In a partial screw shift, XS, only part of the nanoparticle undergoes the relative displacement, resulting in a small twist between the two grains. We measure this twist using the angle between reference atomic columns parallel to the tilt axis in the two grains. In P4 (X0 $\rightarrow$ XS $\rightarrow$ X1), a column shift is produced through two successive partial screw shifts. The second screw shift largely reverses the first, except at a single atomic column which accumulates a displacement equal to one column shift. This is clearly seen in the \emph{z}-displacement maps, where the screw displacement is reversed everywhere except at this column. In P5 (X0 $\rightarrow$ XS $\rightarrow$ X0), the partial screw shift is completely reversed. Finally, in P6 (X0 $\rightarrow$ XS $\rightarrow$ XF), two complementary partial screw shifts combine to produce a full screw shift.We provide two simulations showing the dynamical sequence of these GB processes in Supplementary Figures \ref{sup_fgr:gb_processes_run5} and \ref{sup_fgr:gb_processes_run18}. These processes are not limited to a single local GB geometry and are observed on symmetric, asymmetric, and locally curved GB segments.

In the present simulations, we did not observe either the conventional parallel kink or the surface diffusion induced parallel kink discussed in the previous section. However, we constructed these competing pathways using NEB where relevant. For example, for P2, which involves a column up shift, we also considered a parallel kink pathway and column down shifts through parallel and anti-parallel kinks. The barriers associated with P2, P4, and P6, together with the competing pathways, are listed in Supplementary Table \ref{sup_tab:barrier_gb_process}. We also attempted to construct a direct X0 $\rightarrow$ XF path using NEB, but the path consistently relaxed to X0 $\rightarrow$ XS $\rightarrow$ XF, showing that the two-step route is preferred within the NEB calculation. Figure \ref{fgr:gb_processes} also shows the statistics of the six GB processes counted over the first 50 simulations. P3 is the most frequent process, while P4 is the least frequent. In addition to the six processes, we observed surface-atom diffusion events and two ambiguous (amb.) events in which an atom from a subsurface column diffuses to a neighboring surface column and leaves behind a vacancy. Both ambiguous events quickly reverse.

The six GB processes observed here can therefore be described using different occurrences and combinations of C and S. In particular, the same X0 to X1 transition can occur through P2, P3, or P4. P2 and P3 involve C directly through C\textsubscript{K-} and C, respectively, whereas P4 combines two screw shifts (S+S) to produce the same net column shift. Thus, the same initial and final configurations can be connected through different atomic pathways. The processes identified here are characteristic of the particle and GB configuration considered and are not intended to form an exhaustive list. Other particle sizes or GB configurations may allow additional processes.

\subsection{Common atomic displacements in $\Sigma3$ and $\Sigma11$ grain boundary evolution}
Here we demonstrate that the C and S atomic displacements govern GB evolution in $\Sigma3$, $\Sigma11$, and their junctions. Supplementary Figure \ref{sup_fgr:atomic_displacements_s3} shows the evolution of a $\Sigma3$ GB containing a \emph{dd}, observed during coalescence of two randomly oriented Pt$_{405}$ nanoparticles in the time window from 130 ps to 860 ps. Overall, the \emph{dd} migrates out through column shifts, leaving a symmetric $\Sigma3$ GB at the end. The column shifts occur either through anti-parallel kinks or as full shifts. We also observe transitions between the $\Sigma3$ GB and an \emph{sd} configuration through screw shifts along a $\langle112\rangle$ direction. These results confirm that the C and S atomic displacements also occur dynamically in a $\Sigma3$ GB.

\begin{figure}[!t]
\centering
  \includegraphics[width=1.00\textwidth]{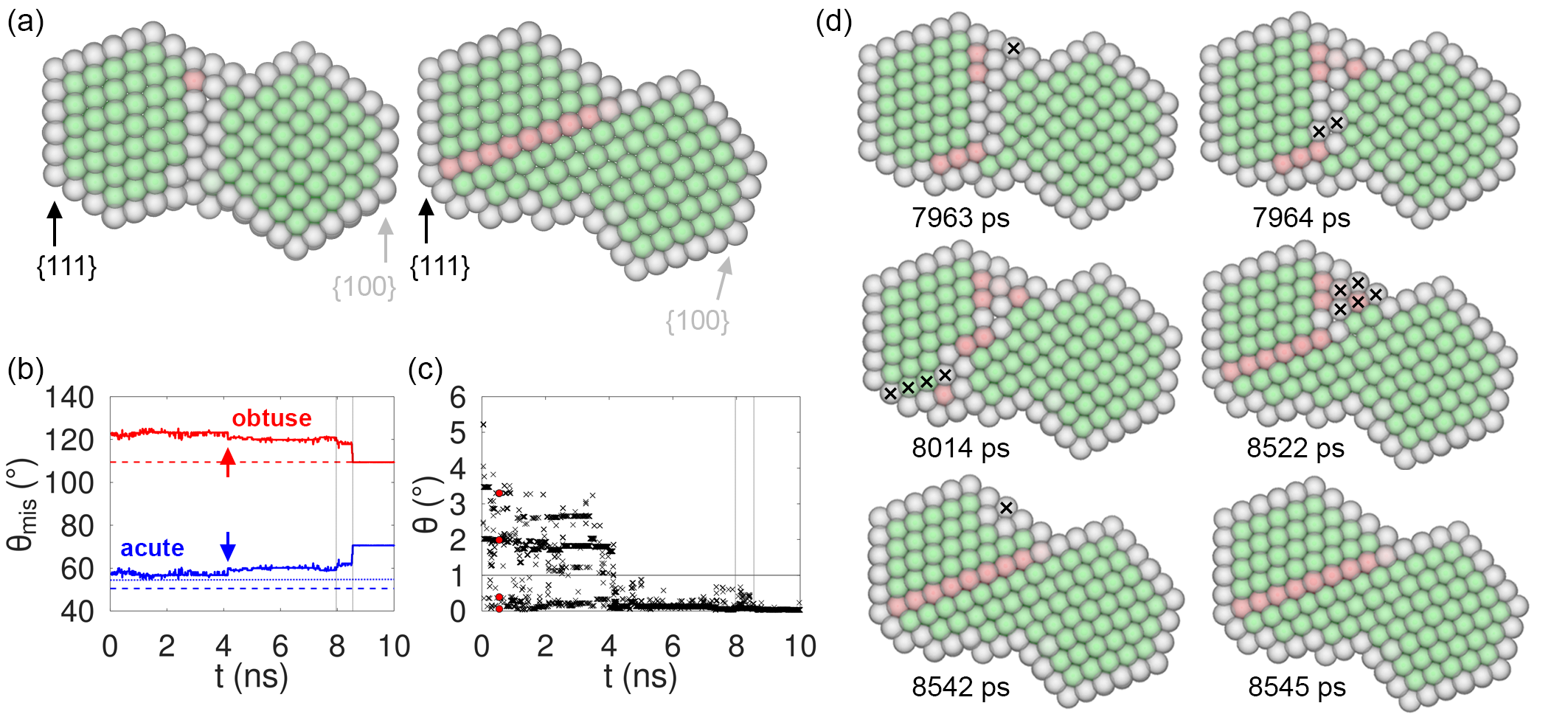}
  \caption{(a) Configurations showing the (left) nearly asymmetric $\Sigma11$ GB and (right) $\Sigma3$ GB (CTB). (b) Misorientation angle (both acute and obtuse) between reference \{100\} planes ($\theta_{mis}$) during the GB transformation. The blue dashed and dotted lines indicate the ideal misorientation angles of symmetric and asymmetric $\Sigma11$ GBs, respectively. The red dashed line indicates the ideal misorientation angle of the symmetric $\Sigma3$ GB. (c) Twist angle ($\theta$) during the GB transformation. The red points correspond to a local change involving screw shifts, which is shown in Supplementary Figure~\ref{sup_fgr:zoom_rand_r13}. (d) Sequence of configurations within the time window between the two vertical lines in panel (c), showing the transformation to the $\Sigma3$ GB through column shifts (indicated by $\times$).}
  \label{fgr:s11_to_s3}
\end{figure}

The compatibility of the GB structural units and the common atomic displacements also makes transformation between $\Sigma11$ and $\Sigma3$ geometrically feasible. Supplementary Figure \ref{sup_fgr:cyl_s11_to_s3} shows the transformation of a symmetric $\Sigma11$ GB to a symmetric $\Sigma3$ GB through suitably chosen column shifts in a cylindrical model. The obtuse angle between reference \{100\} planes on either side of the GB is $\sim130^\circ$ for the symmetric $\Sigma11$ GB, corresponding to a misorientation of $\sim50^\circ$. During transformation to the $\Sigma3$ GB, this angle changes to $\sim110^\circ$, requiring a reduction of $\sim20^\circ$. The local rotation field shows opposite rotations in the regions forming the final $\Sigma3$ configuration, which accommodate this change in orientation. Thus, the transformation requires significant rotational accommodation within the particle, which is possible in unconstrained systems such as nanoparticles.

We observe a similar transformation during coalescence of randomly oriented Pt$_{405}$ nanoparticles. After 40 ps, the two particles develop a GB with an orientation close to an asymmetric $\Sigma11$ (left image in Figure \ref{fgr:s11_to_s3}a), which eventually transforms to a symmetric $\Sigma3$ GB (right image in Figure \ref{fgr:s11_to_s3}a). The rotational accommodation is reflected by the change in angle between the initially approximately parallel \{111\} and \{100\} surface facets, indicated by the black and gray arrows, respectively. We track the transformation using the acute and obtuse angles between two reference \{100\} planes, denoted $\theta_{mis}$ (Figure \ref{fgr:s11_to_s3}b), together with the twist angle (Figure \ref{fgr:s11_to_s3}c). Initially, $\theta_{mis}$ is close to that of an asymmetric $\Sigma11$ GB (blue dotted line). The twist angle shows that partial screw shifts are active up to $\sim4$ ns, beyond which only column shifts are observed ($\theta<1^\circ$). A sequence of partial screw and column shifts corresponding to the red points marked in Figure \ref{fgr:s11_to_s3}c is shown in Supplementary Figure \ref{sup_fgr:zoom_rand_r13}. The first significant change in $\theta_{mis}$ occurs just after $\sim4$ ns. As detailed in Supplementary Figure \ref{sup_fgr:mis_step_rand_r13}, atoms in a surface \{111\} facet shift along a $\langle112\rangle$ directions to form a subsurface $\Sigma3$ GB, followed by a column shift.

The next major change in $\theta_{mis}$ occurs in the time window marked by the two vertical lines beginning at $\sim8$ ns, during which the transformation to $\Sigma3$ proceeds through column shifts alone. Figure \ref{fgr:s11_to_s3}d shows the structural changes, with the atomic columns that undergo stacking change to produce the next configuration marked by $\times$. At 7963 ps, the GB consists of symmetric and asymmetric $\Sigma11$ segments, with a $\Sigma3-\Sigma11$ junction disclination at the bottom and a $\Sigma3$ segment emanating from the junction between the symmetric and asymmetric $\Sigma11$ segments at the top (Supplementary Figure \ref{sup_fgr:junction_defects}a). After three column shifts, the configuration at 8014 ps consists of a $\Sigma3$ GB with a \emph{dd} at the bottom and a faulted \emph{sd} at the top (Supplementary Figure \ref{sup_fgr:junction_defects}b). A further four column shifts migrate the \emph{dd} out of the particle, leaving the faulted \emph{sd} at 8522 ps. Another set of column shifts removes the SF and forms the final $\Sigma3$ CTB at 8545 ps, which is retained until the end of the simulation.

We present three additional examples in the Supplementary Information: a $\Sigma3-\Sigma11$ double junction disclination transforming to a $\Sigma3$ GB (Supplementary Figure \ref{sup_fgr:DJ_to_s3}); a faceted $\Sigma11$ transforming through a $\Sigma3-\Sigma11$ junction and finally to $\Sigma3$ (Supplementary Figure \ref{sup_fgr:DJ_to_s3_s11_migration_r49}); and a $\Sigma3-\Sigma11$ junction impinging on a $\Sigma3$ GB to form a $\Sigma3-\Sigma3-\Sigma11$ triple junction disclination (Supplementary Figure \ref{sup_fgr:DJ_hits_s3_to_TJ_r23}). In the first case, in addition to the rotational accommodation, we observe relative translation of the two lattice regions along the common $\langle110\rangle$ axis. Changes in this relative translation occur approximately in multiples of the interplanar distance of the \{110\} planes (Supplementary Figure \ref{sup_fgr:zdiff_r13_r72}b). In contrast, there is no net relative translation along the tilt axis for the transformation in Figure \ref{fgr:s11_to_s3} (Supplementary Figure \ref{sup_fgr:zdiff_r13_r72}a). Thus, rotational accommodation is a common feature of the $\Sigma11$ to $\Sigma3$ transformations considered here, while relative translation along the common $\langle110\rangle$ axis provides an additional degree of freedom that may or may not occur.

%Across all these examples, the atomic changes that we resolve are described by the same C and S displacements introduced in Figure \ref{fgr:disconnection_struct}. The specific GB processes vary with the local structure: column shifts can occur through kinks or as full shifts, while screw shifts can reverse, combine to produce a full regional shift, or combine to leave a net column shift. Thus, $\Sigma3$, $\Sigma11$, and their junctions can undergo different GB processes while using the same fundamental atomic displacements.

Across these examples, the C and S displacements remain identifiable as the local GB structure evolves between $\Sigma11$, $\Sigma3$, and their junction configurations. This includes transformations involving significant rotational accommodation and changes in the GB character itself. Thus, although the detailed GB processes depend on the local structure, the same atomic displacement description remains useful throughout the evolution.

\section{Discussion and Summary}
\label{sec:conclusions}

A clear picture emerges from the results presented here for the evolution of low energy GBs in nanoparticles. From a structural point of view, GB evolution exhibits a rich landscape involving symmetric, asymmetric, faceted, and locally curved boundaries, transformations between $\Sigma11$ and $\Sigma3$, and evolution of their junctions. However, this structural diversity can be described using the fundamental atomic displacements C and S. The complexity then arises from how these displacements occur and combine under dynamical conditions. The same structural change can be achieved through different atomic pathways, while different combinations of C and S can result in migration, GB transformation, or junction evolution. These displacements remain identifiable even when the GB character itself changes during the transformation.

GB evolution can be described using the disconnection framework and also through the underlying atomic displacements, often referred to as atomic shuffles. Prior works have reported C type displacements of CTPs in $\Sigma11$ GBs \cite{wang2026_Pt_GBM} and incoherent twin boundaries (ITBs) \cite{brown2009_CTP_ITB}, where CTPs with the same stacking are separated by SFs. Kinked intermediate configurations and kinked disconnection migration have also been reported \cite{combe2016_disconnection_kinks,larranaga2020_sessile_disconnections}. The disconnection framework is most naturally applicable when the GB character remains well defined. However, its application becomes difficult when the GB undergoes dynamical structural changes \cite{combe2021_multiple_coupling_modes}. Similar transformations have been observed in Au nano-bicrystals, where a faceted GB involving $\Sigma11$ structures evolves to a symmetric $\Sigma11$ GB or develops a $\Sigma3$ GB during the evolution \cite{fang2022_GB_transformation}. In our simulations, during the $\Sigma11$ to $\Sigma3$ transformation, the misorientation itself changes, and therefore the same coincident site lattice (CSL) and displacement shift complete (DSC) lattice cannot be retained throughout the transformation. Under this scenario, tracking the complete GB evolution using a disconnection description becomes cumbersome. However, since the same C and S displacements occur across $\Sigma11$, $\Sigma3$, and the intermediate configurations observed here, the complete evolution can be described using a common set of atomic displacements.

Different sources of complexity in GB evolution have been identified in previous studies. Multiple stable or metastable configurations have been observed during GB evolution \cite{wei2021_atomistic_GBM,frolov2013_GB_phases}. For a given GB type, multiple disconnection modes, corresponding to different combinations of Burgers vector and step height, can also be energetically competitive \cite{han2018_disconnection,combe2021_multiple_coupling_modes,gautier2025_GB_deformation}. During GB evolution, these modes can switch dynamically \cite{thomas2017_grain_growth_GBM} and can combine or decompose to produce other disconnection modes \cite{rajabzadeh2014_disconnection_GBM}. For example, two single layer disconnections in a $\Sigma11$ Au GB can combine to form a double layer disconnection \cite{zhu2019_Au_disconnection}. Our results show another level of complexity at the atomic pathway level. Even when the initial and final compact structural configurations are the same, the atomic path connecting them is not unique. For a column shift X0 $\rightarrow$ X1, the C displacement can occur through a full column shift or through kinked pathways C\textsubscript{K-} and C\textsubscript{K+}. More importantly, the same X0 $\rightarrow$ X1 transition can also be obtained through two successive S displacements, whose combined effect produces the same net C displacement (S + S $\rightarrow$ C). Thus, the path can differ not only in how a given displacement occurs, but also in which fundamental atomic displacements are combined to reach the same final configuration. The S + S $\rightarrow$ C process is related in spirit to previously reported reactions between disconnections, but here the combination is described at the atomic displacement level as an alternative atomic path to the same structural change. More generally, different combinations of C and S also give rise to different GB processes, even when they do not connect the same initial and final configurations.

Due to their finite size, nanoparticles can accommodate structural changes involving rotational and translational degrees of freedom, as evident from the $\Sigma11$ to $\Sigma3$ transformations observed in our simulations. Compared to GBs in polycrystalline materials, GBs in nanoparticles also terminate at free surfaces, which naturally provide heterogeneous nucleation sites for disconnection kinks. In all the kinked column shifts observed in our simulations and NEB calculations, the kink nucleates at the surface and then propagates through the column. In addition, we observed that surface diffusion of an atom from a neighboring column can trigger a kinked column shift. These results suggest that surfaces can play an important role in determining the GB processes available in nanoparticles.

The finite column length also affects the competition between full and kinked column shifts. We compared their energy barriers as a function of column size for $\Sigma11$ and $\Sigma3$ GBs in nanoparticles and bicrystals using a \emph{frozen-z} method, which provides a controlled comparison of the prescribed pathways but does not necessarily correspond to the minimum energy path (Supplementary Figure \ref{sup_fgr:barrier_flattening}a). The barrier for a full shift increases approximately linearly with the number of atoms in the column, whereas the anti-parallel kink barrier tends to plateau, with the two becoming comparable for very short columns. NEB calculations show the same qualitative plateauing behavior, and for the cases examined the parallel kink has a higher barrier than the anti-parallel kink (Supplementary Figure \ref{sup_fgr:barrier_flattening}b). Thus, as the column length increases, a localized kinked pathway becomes increasingly favorable compared with a concerted full shift.

Identifying the active GB process also provides a way to target the local chemistry and modify its kinetics. We tested this idea in an idealized Cu bicrystal containing a symmetric $\Sigma11$ GB. Substituting Ag into the Cu column adjacent to a \emph{dd} increases the energy required for the column shift on average and also changes the shear displacement required to migrate the \emph{dd} (Supplementary Figure \ref{sup_fgr:barrier_modification}). This suggests that identifying the atomic pathway can provide a route to modify GB migration through local chemical changes.

The present results may also be relevant to a broader range of nanoparticle structures. CTBs are intrinsic to twinned, decahedral, and icosahedral nanoparticles \cite{song2020_fivefold_twins,song2024_detwinning,song2024_pentatwinned_Au,sun2025_icosahedral_AuPd}, while low energy special GBs are also frequently observed in nanoparticle assemblies \cite{geng2024_nano_assemblies,zhu2022_GBCD_AuNP}. However, GB populations in nanoparticles are commonly classified based on grain misorientation, which does not necessarily specify the local atomic GB structure in a finite particle \cite{settem2024_GBJ_disclinations}. The evolution of these structures has been described through grain rotation, disconnection or partial-dislocation motion, surface diffusion, and structural transformation. These systems therefore provide natural cases to test whether the C and S description extends beyond the compact $\Sigma3$ and $\Sigma11$ configurations studied here. This may also be relevant in reactive environments, where surface chemistry can drive twin boundary migration \cite{carnis2021_CO_TB_migration} and could also alter the atomic pathway itself.

The six GB processes identified for the faceted $\Sigma11$ configuration are not intended to form an exhaustive description of GB evolution. Other particle sizes, GB configurations, and conditions may allow additional processes. Similarly, the C and S displacements have been demonstrated here for compact $\Sigma3$, $\Sigma11$, and their junction configurations, and their applicability to other GB structures remains to be established. Nevertheless, the results show that a variety of GB processes can arise from different occurrences and combinations of C and S. The same structural change can also proceed through different atomic pathways, showing that the initial and final GB configurations alone do not uniquely determine how the transformation occurs. Identifying these atomic displacements and their pathways therefore provides a way to understand how local structure, nanoparticle geometry, surfaces, and chemistry influence GB evolution and potentially how it can be controlled.

\section{Methods}
\label{sec:methods}

The interatomic interactions were described using embedded atom method (EAM) potentials \cite{daw1984_EAM}. For Pt, we used the potential of Sheng \emph{et al.} \cite{sheng2011_Pt_EAM}, while the Ag and Cu--Ag calculations were carried out using the potential of Williams \emph{et al.} \cite{williams2006_CuAg_EAM}. These potentials were chosen because the ratio of the $\Sigma11$ and $\Sigma3$ GB energies, $\gamma_{\Sigma11}/\gamma_{\Sigma3}$, obtained using these potentials agrees reasonably well with DFT values available in the literature (see Supplementary Information of Ref.~\cite{settem2024_GBJ_disclinations}). This is important here because the competition between $\Sigma11$ and $\Sigma3$ structures is central to the GB evolution considered in this work. Molecular dynamics (MD) simulations and nudged elastic band (NEB) calculations were performed using LAMMPS \cite{plimpton1995_LAMMPS}.

All Pt and Ag coalescence simulations were carried out at a homologous temperature $T_{\mathrm{hom}}=T/T_{\mathrm{mp}}\sim0.45$, where $T_{\mathrm{mp}}$ is the bulk melting temperature predicted by the corresponding potential. This gives 850~K for Pt ($T_{\mathrm{mp}}\sim1890$~K \cite{sheng2011_Pt_EAM}) and 570~K for Ag ($T_{\mathrm{mp}}\sim1267$~K \cite{williams2006_CuAg_EAM}). The equations of motion were integrated with a timestep of 5~fs, and the temperature was maintained using a Langevin thermostat with a damping time of 0.5~ps. The nanoparticle simulations were carried out with non-periodic boundaries in all three directions. The center of mass was recentered during the simulations, and the overall linear and angular momenta were periodically removed. The homologous temperature was chosen to allow GB evolution within timescales accessible to MD simulations while remaining sufficiently below the melting temperature to avoid extensive disordering.

We considered two types of nanoparticle coalescence simulations. For the randomly oriented simulations, two copies of a 405-atom truncated octahedron were independently rotated by sequential rotations about the $x$, $y$, and $z$ axes, with each rotation angle drawn independently from a uniform distribution between $0^\circ$ and $360^\circ$. The second particle was then translated along the $x$ direction such that $\min(x_2)-\max(x_1)=3$~\AA. We also carried out oriented attachment simulations in which a \{100\} facet of the left particle was attached to a \{111\} facet of the right particle. The particles were oriented such that the two orthogonal $\langle100\rangle$ directions of the left particle were parallel to the orthogonal $\langle110\rangle$ and $\langle112\rangle$ directions of the right particle. This geometry naturally forms a $\Sigma11$ GB and was used for the 711-atom + 807-atom Pt system shown in Figure~\ref{fgr:disconnection_kinks}. The initial configuration is shown at 0~ns in Figure~\ref{fgr:disconnection_kinks}a.

For each metal, we carried out 80 independent 100~ns coalescence simulations of randomly oriented 405-atom truncated octahedra to examine the GB structures retained after coalescence. Different initial relative orientations and random seeds for the initial velocities were used for the independent simulations. The configurations at 10 and 100~ns, giving 160 configurations for each metal, were visually inspected and classified as fcc (defect-free particle), $\Sigma3$ (one CTB), $\Sigma3+\Sigma3$ (two CTBs), SF, $\Sigma11$, GBJ disclination ($\Sigma3-\Sigma3-\Sigma11$ or $\Sigma3-\Sigma11$), $\Sigma11$ + inclined $\Sigma3$, $\Sigma11$ + SF, or Dh.

A separate set of 80 randomly oriented 405-atom nanoparticle coalescence simulations was carried out for 10~ns, with configurations sampled every 1~ps, to follow the GB evolution more closely. These simulations were not used to determine GB structure statistics, but instead to identify representative intermediate GB configurations and examples of GB evolution involving $\Sigma11$, $\Sigma3$, and their junctions. Since a temporal resolution of 1~ps was not sufficient to resolve the individual atomic-scale processes, selected intermediate configurations were used to initiate additional 5 or 10~ps MD simulations, with configurations sampled every 10~fs.

For the system shown in Figure~\ref{fgr:disconnection_kinks}, 70 independent 10~ps simulations were initiated from the intermediate configuration at 29.5~ns using independently initialized velocities, with configurations sampled every 10~fs. For the system shown in Figure~\ref{fgr:gb_processes}, 100 independent 5~ps simulations were initiated from an intermediate GB configuration obtained during coalescence of two randomly oriented 405-atom nanoparticles, again using independently initialized velocities and sampling configurations every 10~fs. The sampled configurations were quenched to 0~K using conjugate-gradient energy minimization in LAMMPS, with energy and force tolerances of $10^{-10}$ and $10^{-10}$~eV/\AA, respectively. Structural changes associated with changes in the quenched energy profile were inspected to identify the six GB processes shown in Figure~\ref{fgr:gb_processes}. Their statistics were obtained by counting the processes occurring across the first 50 simulations.

To identify the fundamental atomic displacements, idealized $\Sigma11$ and $\Sigma3$ bicrystal cylinders with a diameter of approximately 20~\AA\ were constructed. The cylinders were periodic along the cylinder axis and free in the transverse directions; for the $\Sigma11$ model, the cylinder axis was parallel to the $\langle110\rangle$ tilt axis. Column or screw shifts were imposed geometrically on the relevant atomic columns or regions, followed by quenching to 0~K. For successive column shifts, each displacement was applied after quenching the preceding configuration. These constructions were used to establish the C and S displacement motifs, while their dynamical occurrence was examined separately in the finite-temperature nanoparticle simulations.

NEB calculations were performed using the \texttt{quickmin} damped-dynamics minimizer in LAMMPS with a minimization timestep of 5~fs. A spring constant of 30~eV/\AA$^{2}$ was used between neighboring replicas. The energy and force convergence tolerances were $10^{-12}$ and $10^{-6}$~eV/\AA, respectively, with a maximum of $10^{5}$ iterations for both the initial NEB and climbing-image stages. The initial and final configurations were first quenched to 0~K using the same conjugate-gradient minimization procedure described above. For the pathways shown in Figure~\ref{fgr:disconnection_kinks}d, the full column shift, parallel kink, and surface diffusion triggered parallel kink calculations used 21 images, while the anti-parallel kink calculation used 71 images. In each case, the NEB pathway was initialized using only the initial and final configurations.

For the concerted full column shift, additional harmonic restraints were applied between neighboring atoms in the shifting column to suppress vacancy-like separation and prevent the pathway from evolving into a kinked shift. The restraint was applied with a constant equilibrium separation for each neighboring atomic pair, taken as the mean of its separation in the initial and final configurations. A restraint coefficient $K_{\mathrm{col}}=20$~eV/\AA$^{2}$ was used for the full-shift pathway shown in Figure~\ref{fgr:disconnection_kinks}d. The restraint energy was included during the NEB minimization, but after convergence the restraints were removed without further atomic relaxation and the EAM energies of the resulting replicas were used to obtain the reported full-shift energy profile. The dependence of the pathway and barrier on $K_{\mathrm{col}}$, together with further details of the NEB calculations, is provided in the Supplementary Section \ref{sup_sec:kcol_robustness}.

During preparation of this manuscript, the authors used ChatGPT (OpenAI) to assist with language editing and rephrasing. The authors reviewed and edited all resulting text and take full responsibility for the content of the manuscript.

%\section*{Acknowledgments}

%\section*{Funding}
%This work did not receive any funding.g.

\bibliographystyle{unsrt}  
\bibliography{arxiv_ms/refs}

\clearpage

\begin{center}
    {\LARGE\bfseries Supplementary Information}
\end{center}

\vspace{1em}

% Supplementary section numbering
\setcounter{section}{0}
\renewcommand{\thesection}{S\arabic{section}}

% Supplementary figure and table numbering
\setcounter{figure}{0}
\setcounter{table}{0}
\renewcommand{\thefigure}{S\arabic{figure}}
\renewcommand{\thetable}{S\arabic{table}}

% Distinct hyperlink anchors for Supplementary Information
\renewcommand{\theHsection}{S\arabic{section}}
\renewcommand{\theHfigure}{S\arabic{figure}}
\renewcommand{\theHtable}{S\arabic{table}}

\section{GB structures obtained from coalescence}
\label{sup_sec:coalescence_structures}

Through coalescence of randomly oriented 405-atom nanoparticles, we assessed the types of GB structures that persist after the transient rearrangements following attachment. Supplementary Table~\ref{sup_tab:disclination_fraction2} lists the number of configurations belonging to each structural class, as defined in the Methods section, at 10 and 100~ns. Representative structures from these classes are shown in Supplementary Figure~\ref{sup_fgr:rand_coal_structs}. Among the GB-containing configurations, the persistent GB structures are composed of $\Sigma3$, $\Sigma11$, or combinations and junctions involving these two GB types. This is consistent with the preference for compact structural-unit GBs at the nanoscale \cite{settem2024_GBJ_disclinations}.

The relative prevalence of $\Sigma3$ and $\Sigma11$ differs considerably between Pt and Ag. At 100~ns, configurations containing only $\Sigma3$-type GBs ($\Sigma3$, $\Sigma3+\Sigma3$, and Dh) account for 28.75\% of Pt and 67.50\% of Ag configurations. Configurations containing only $\Sigma11$-type GBs ($\Sigma11$ and $\Sigma11+$ SF) account for 32.50\% of Pt and 1.25\% of Ag configurations, while configurations containing both $\Sigma3$ and $\Sigma11$ account for 7.50\% and 6.25\%, respectively. The greater prevalence of $\Sigma11$-containing structures in Pt is consistent with its lower $\gamma_{\Sigma11}/\gamma_{\Sigma3}$ ratio compared with Ag.

\begin{table*}[!ht]
\footnotesize
\centering
\caption{Structural distribution following coalescence of two randomly oriented 405-atom fcc nanoparticles. The number of configurations belonging to each structural class, out of 80 independent simulations for each metal, is reported at 10 and 100~ns. SF $\rightarrow$ stacking fault, GBJ $\rightarrow$ grain boundary junction, and Dh $\rightarrow$ decahedron.}
{
\def\arraystretch{1.5}
\begin{tabular*}{1.00\textwidth}{@{\extracolsep{\fill}}l|cc|cc}
\hline
\hline
\textbf{Structure} & \multicolumn{2}{c|}{\textbf{Pt (850 K)}} & \multicolumn{2}{c}{\textbf{Ag (570 K)}}\\
{} & {10 ns} & {100 ns} & {10 ns} & {100 ns}\\
\hline
{fcc} & {23} & {24} & {11} & {5}\\
{$\Sigma3$} & {20} & {22} & {26} & {26}\\
{$\Sigma3+\Sigma3$} & {0} & {1} & {17} & {20}\\
{SF} & {1} & {1} & {11} & {15}\\
{$\Sigma11$} & {21} & {24} & {1} & {0}\\
{GBJ disclination} & {10} & {5} & {6} & {5}\\
{$\Sigma11+$ inclined $\Sigma3$} & {2} & {1} & {1} & {0}\\
{$\Sigma11+$ SF} & {3} & {2} & {4} & {1}\\
{Dh} & {0} & {0} & {3} & {8}\\
\hline
\hline
\end{tabular*}
}
\label{sup_tab:disclination_fraction2}
\end{table*}

\begin{figure}[!h]
\centering
  \includegraphics[width=1.0\textwidth]{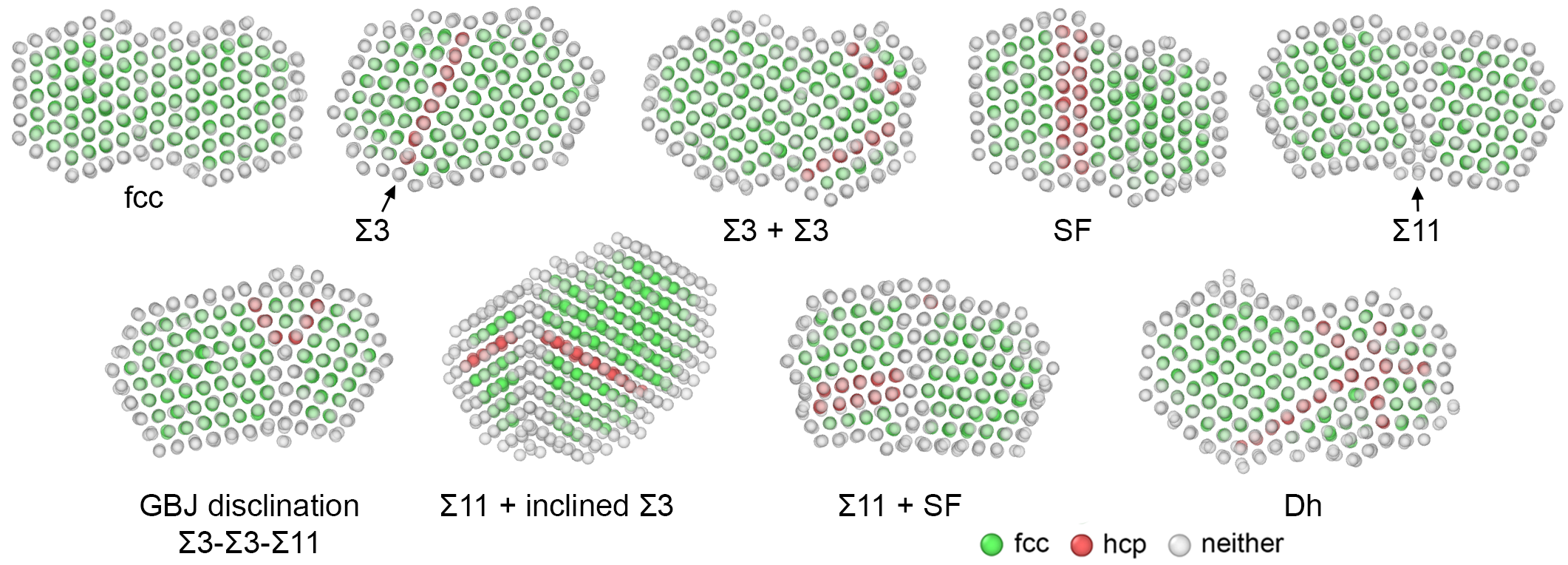}
  \caption{Representative structural classes observed following nanoparticle coalescence. All configurations shown are Pt nanoparticles at 100~ns, except for the Dh structure, which is shown for Ag.}
  \label{sup_fgr:rand_coal_structs}
\end{figure}

\section{Structural units of symmetric and asymmetric $\Sigma11$ GBs}

CTPs define the structure of both symmetric and asymmetric $\Sigma11$ segments. In Supplementary Figure~\ref{sup_fgr:as11_displ}a, column shifts in the lower region (indicated by red columns in the $z$-displacement map), with respect to the initially symmetric $\Sigma11$ GB, transform it into a faceted GB consisting of symmetric and asymmetric $\Sigma11$ segments. Additional column shifts in the upper region further transform it into a predominantly asymmetric $\Sigma11$ GB (Supplementary Figure~\ref{sup_fgr:as11_displ}b). A symmetric $\Sigma11$ segment consists of staggered CTPs with alternating stacking, indicated by blue and orange triangles. On the other hand, an asymmetric $\Sigma11$ segment consists of CTPs with the same stacking, in this case indicated by orange triangles. Thus, both symmetric and asymmetric $\Sigma11$ GBs are defined by CTPs, with the asymmetric $\Sigma11$ consisting of only one type of CTP.

\begin{figure}[!h]
\centering
  \includegraphics[width=1.0\textwidth]{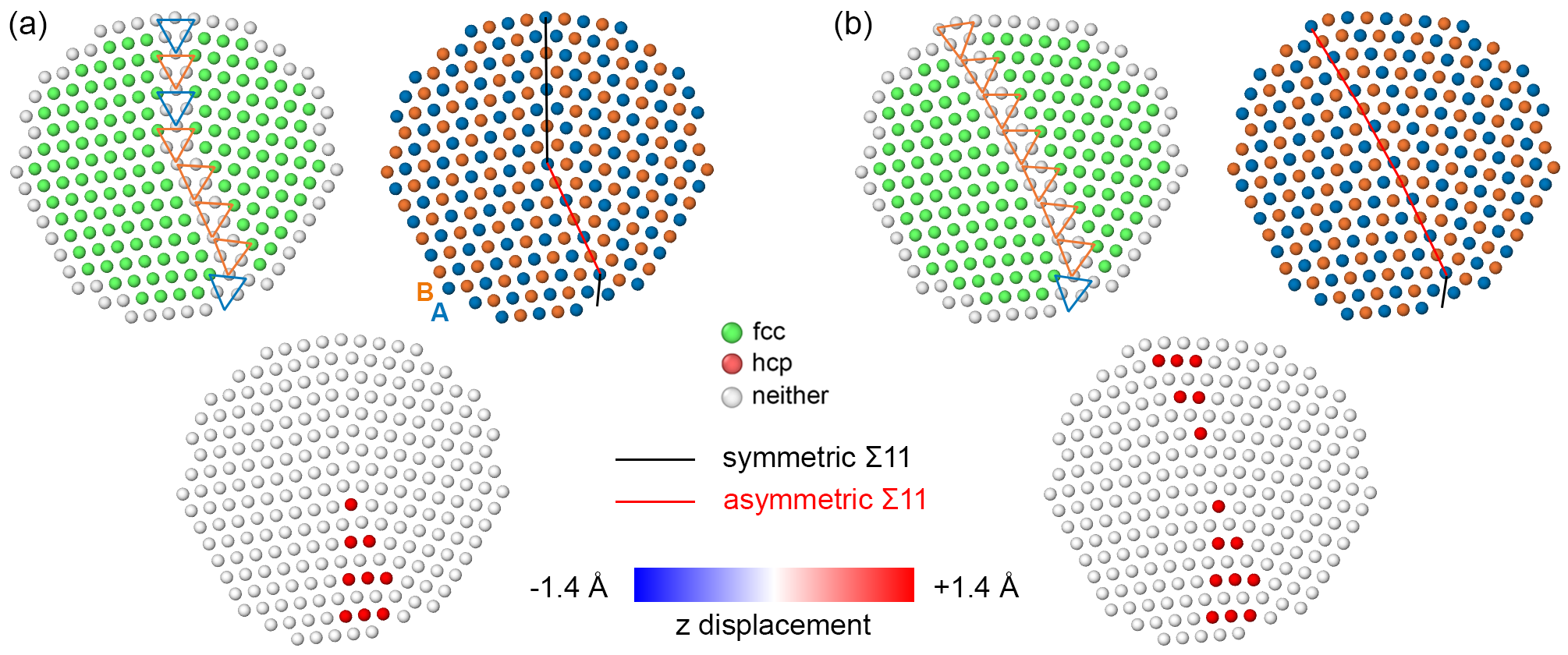}
  \caption{Column shifts that transform an initially symmetric $\Sigma11$ GB into configurations containing symmetric and asymmetric $\Sigma11$ segments. The columns undergoing stacking change are shown in red in the $z$-displacement maps.}
  \label{sup_fgr:as11_displ}
\end{figure}

\section{Structure and energetics of $\Sigma11$ disconnections}

\begin{figure}[!h]
\centering
  \includegraphics[width=1.0\textwidth]{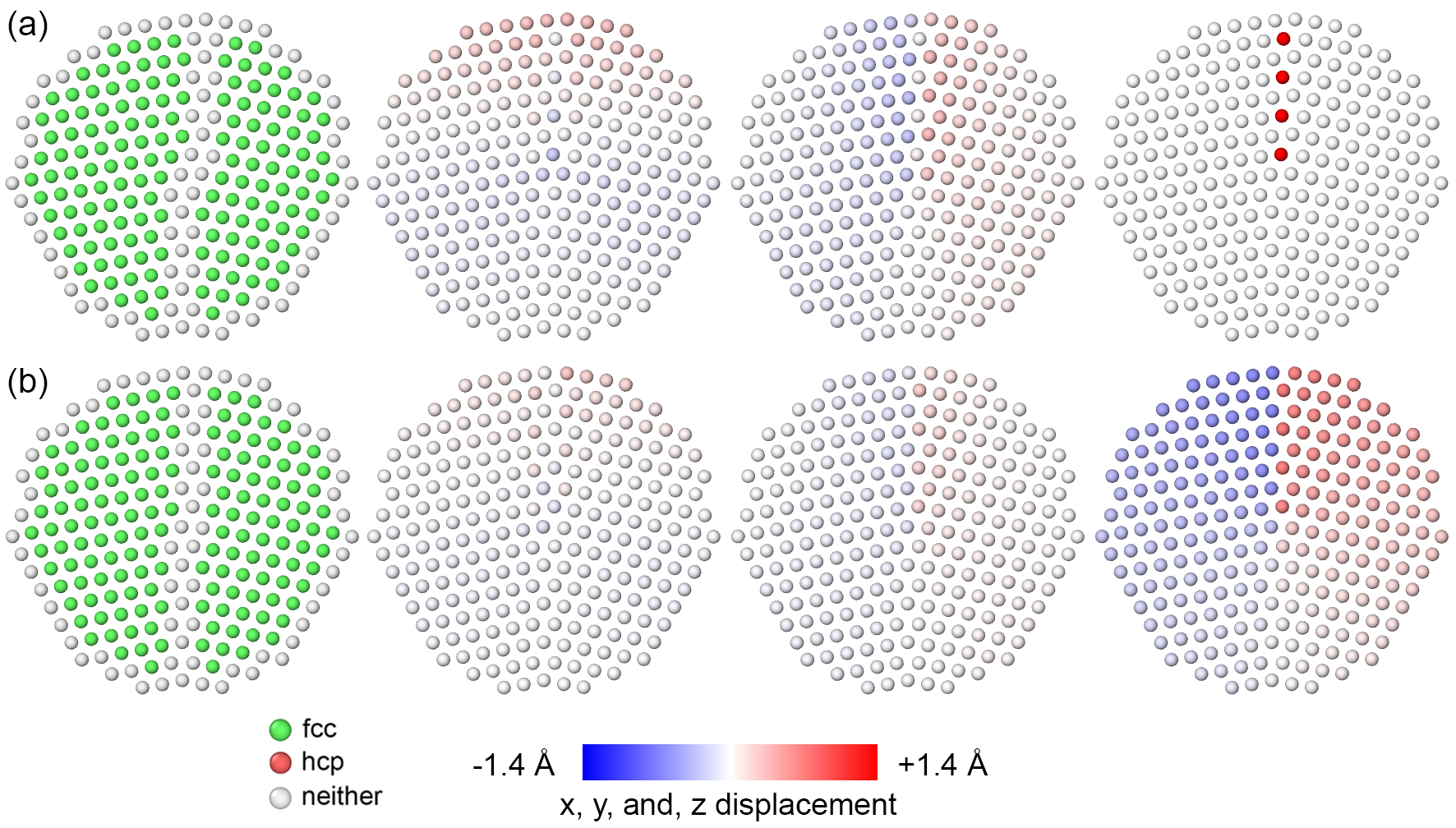}
  \caption{Structure and displacement maps of (a) a double disconnection (\emph{dd}) and (b) a single disconnection (\emph{sd}) in a $\Sigma11$ GB. From left to right: disconnection structure, $x$-displacement map, $y$-displacement map, and $z$-displacement map.}
  \label{sup_fgr:sd_dd_displ_maps}
\end{figure}

The $x$-, $y$-, and $z$-displacement maps of the double disconnection (\emph{dd}, Supplementary Figure~\ref{sup_fgr:sd_dd_displ_maps}a) and single disconnection (\emph{sd}, Supplementary Figure~\ref{sup_fgr:sd_dd_displ_maps}b) show that the displacement in both cases occurs predominantly in the $z$ direction, i.e., along the tilt axis, while the in-plane atomic rearrangements are comparatively smaller. In addition, the \emph{sd} has a larger spatial extent of displacement compared with the \emph{dd}, for which the column shifts are highly localized.

\begin{figure}[!h]
\centering
  \includegraphics[width=1.0\textwidth]{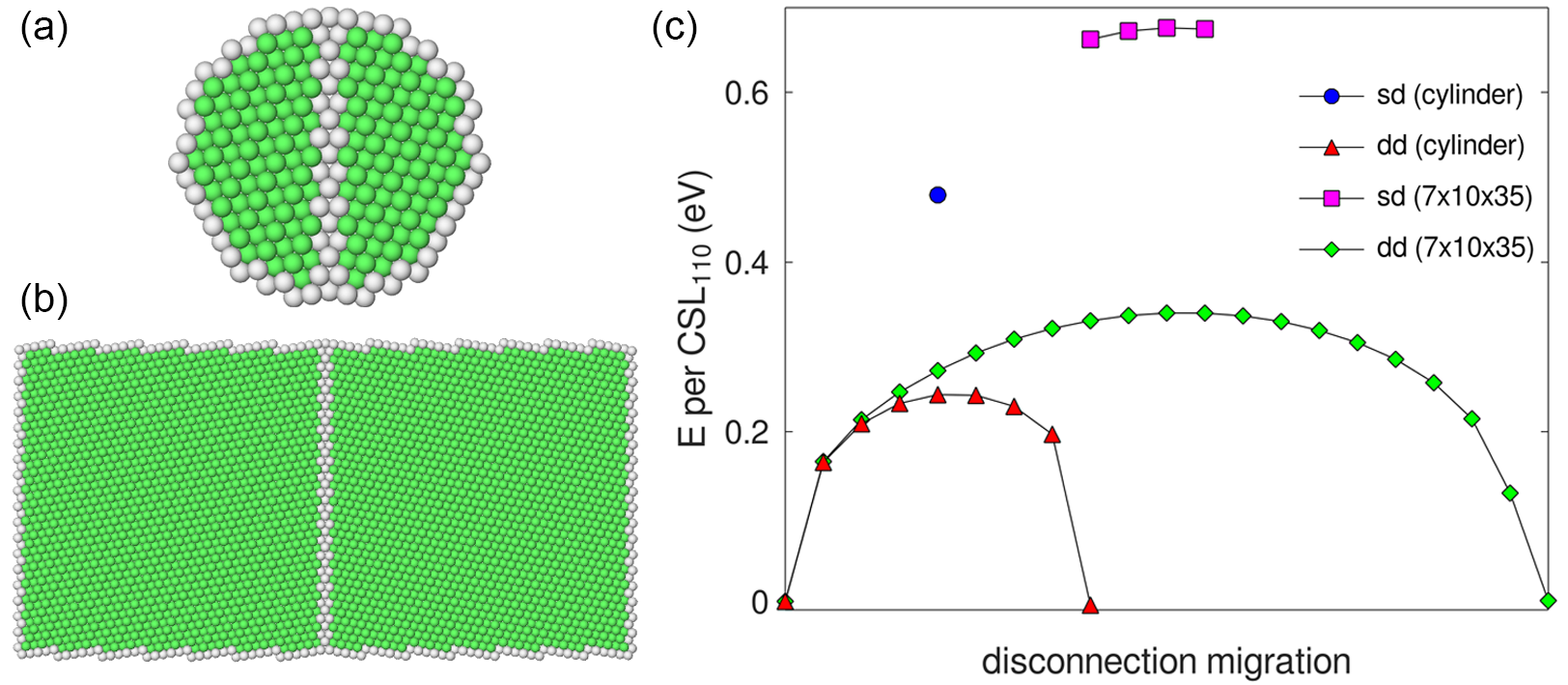}
  \caption{Symmetric $\Sigma11$ GB models used to compare the energetics of single and double disconnections: (a) a 2~nm diameter cylinder and (b) a rectangular bicrystal in which each grain has dimensions of 7, 10, and 35 CSL units along $\langle113\rangle$ ($x$), $\langle332\rangle$ ($y$), and $\langle110\rangle$ ($z$), respectively. (c) Energies of single disconnection (\emph{sd}) and double disconnection (\emph{dd}) configurations in the cylinder and rectangular bicrystal. Both models are periodic only along $\langle110\rangle$ ($z$). The two endpoints of the \emph{dd} sequence correspond to disconnection-free configurations.}
  \label{sup_fgr:discon_energetics}
\end{figure}

We calculated the energies of different \emph{dd} configurations as the disconnection migrates between two neighboring symmetric $\Sigma11$ GB positions using two models: a 2~nm diameter cylinder (Supplementary Figure~\ref{sup_fgr:discon_energetics}a) and a bicrystal (Supplementary Figure~\ref{sup_fgr:discon_energetics}b) in which each grain has dimensions of 7, 10, and 35 CSL units along $\langle113\rangle$, $\langle332\rangle$, and $\langle110\rangle$, respectively. Both models are periodic along the $\langle110\rangle$ tilt axis and finite in the other two directions. We also calculated the energies of corresponding \emph{sd} configurations at similar locations along the initial $\Sigma11$ GB. Only \emph{sd} configurations near the central region of the GB remained stable after quenching to 0~K, while the others reverted to configurations without the imposed \emph{sd}. For the stable configurations, the \emph{sd} has a higher energy than the \emph{dd} at comparable locations along the GB (Supplementary Figure~\ref{sup_fgr:discon_energetics}c).  

\clearpage
\section{Characteristics of $\Sigma11$ and $\Sigma3$ disconnections}

\begin{figure}[!h]
\centering
  \includegraphics[width=0.77\textwidth]{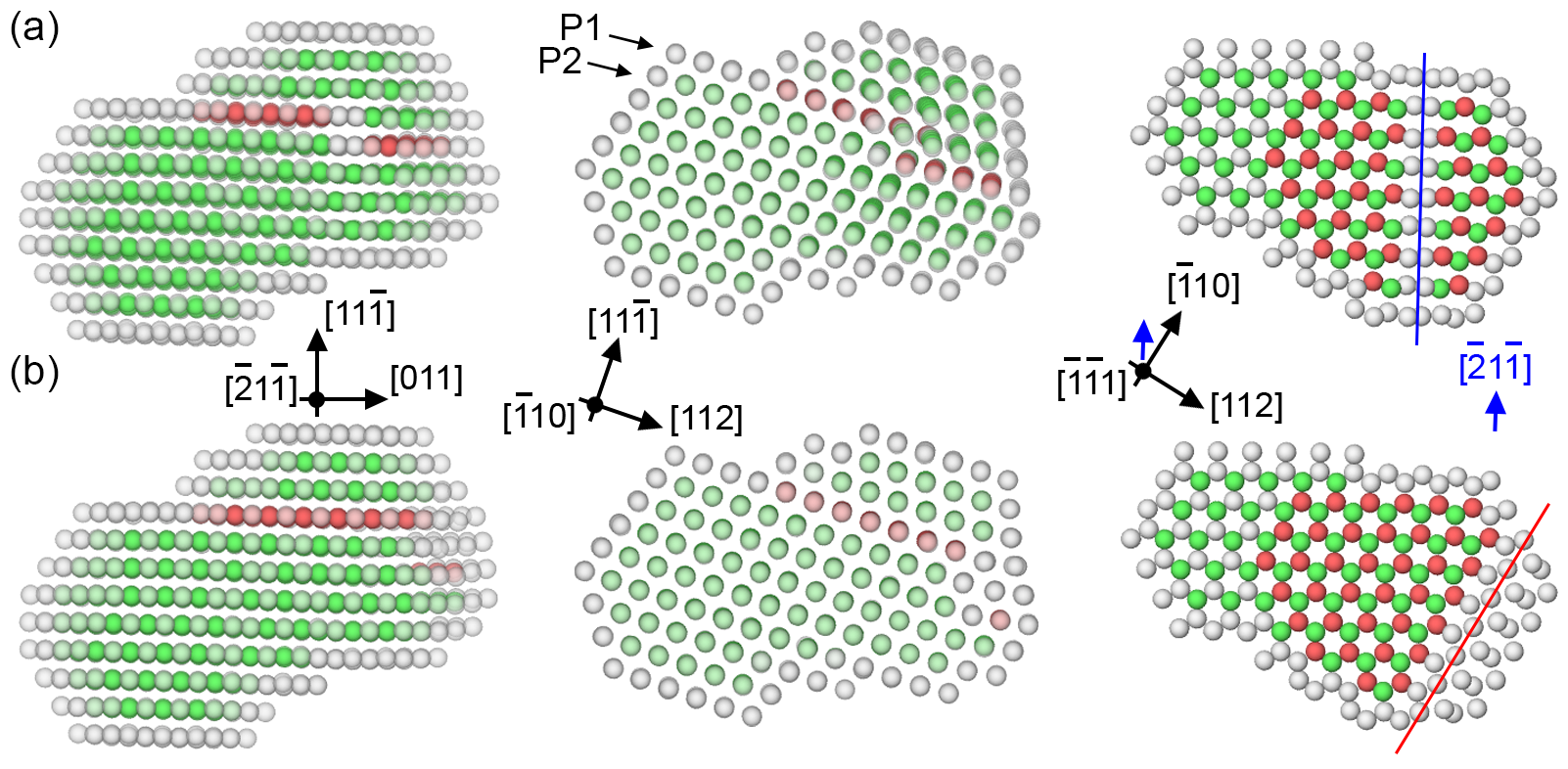}
  \caption{(a) \emph{sd} and (b) \emph{dd} configurations of a $\Sigma3$ GB observed during coalescence of two randomly oriented 405-atom nanoparticles, shown in three projections (from left to right): $[\bar{2}1\bar{1}]$, $[\bar{1}10]$, and $[\bar{1}\bar{1}1]$. In the $[\bar{1}\bar{1}1]$ projection, only the \{111\} planes P1 and P2 are shown. The disconnection lines of the \emph{sd} (blue) and \emph{dd} (red) are also indicated.}
  \label{sup_fgr:s3_sd_dd_disconn_lines}
\end{figure}

\begin{figure}[!h]
\centering
  \includegraphics[width=1.0\textwidth]{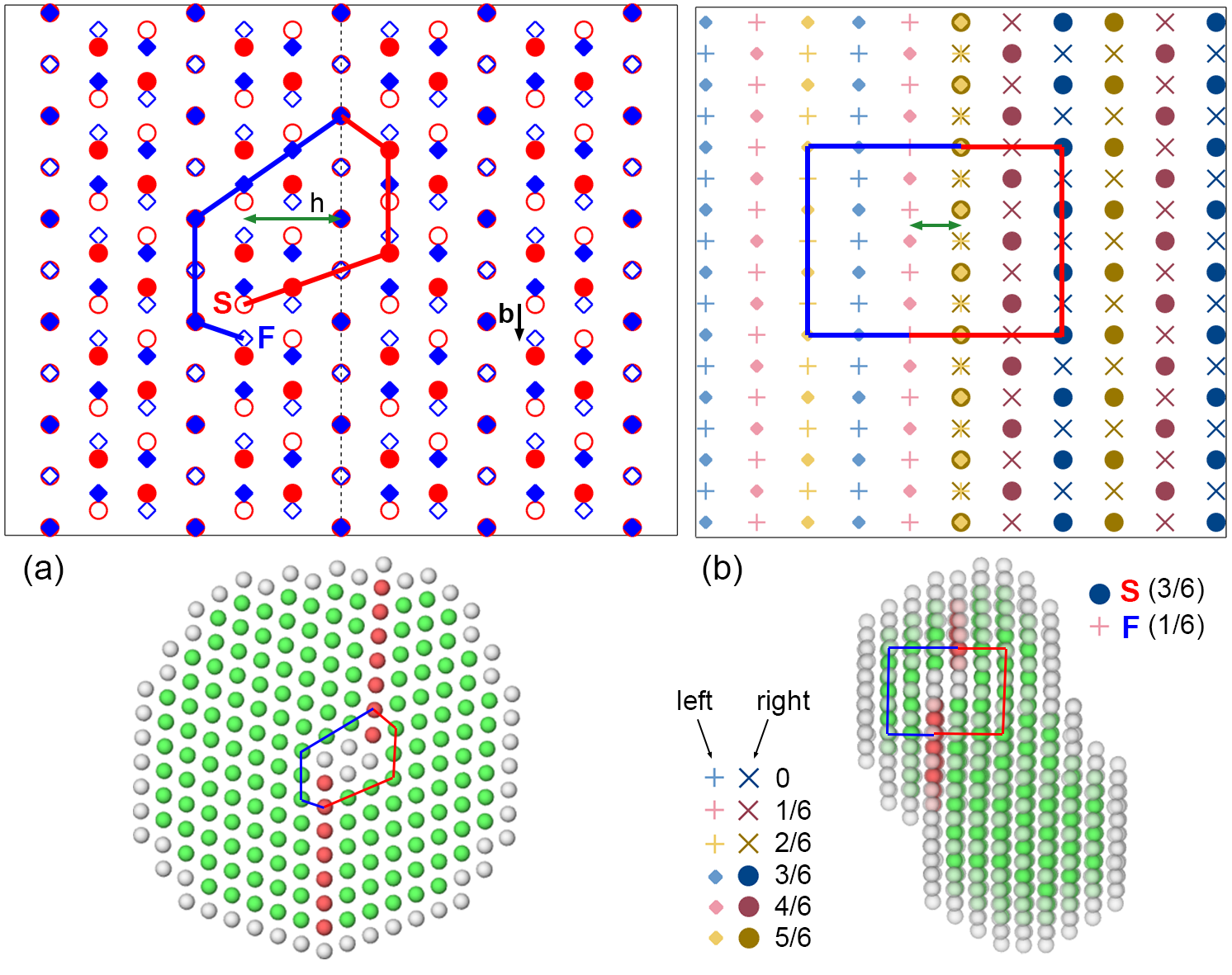}
  \caption{Burgers circuits mapped onto the dichromatic pattern for (a) a $\Sigma3$ double disconnection (\emph{dd}), viewed along $\langle110\rangle$, and (b) a $\Sigma3$ single disconnection (\emph{sd}), viewed along $\langle112\rangle$. In the dichromatic pattern in (a), the right and left crystals are shown by circles and diamonds, respectively. Open and solid symbols refer to heights $z=0$ and $z=+d_{110}$, respectively, where $d_{110}$ is the distance between \{110\} planes. In the dichromatic pattern in (b), the right and left crystals and the atomic heights are indicated by the marker legend. For clarity, the right and left crystals are shown only for $x\geq0$ and $x\leq0$, respectively. The heights are given as fractions of the CSL repeat distance along $\langle112\rangle$, which is equal to the magnitude of $\frac{1}{2}\langle112\rangle$. \emph{S} and \emph{F} indicate the start and finish of the Burgers circuit, respectively. $\boldsymbol{b}$ and $h$ are the Burgers vector and step height, respectively. The displacement-shift-complete (DSC) vectors of $\Sigma3$ are $\boldsymbol{D}_x=\frac{1}{3}[\bar{1}\bar{1}1]$, $\boldsymbol{D}_y=\frac{1}{6}[112]$, and $\boldsymbol{D}_z=\frac{1}{4}[\bar{1}10]$. The Burgers vectors of the double disconnection and single disconnection are $\boldsymbol{b}_{dd,\Sigma3}=-\boldsymbol{D}_y=\frac{1}{6}[\bar{1}\bar{1}\bar{2}]$ and $\boldsymbol{b}_{sd,\Sigma3}=\frac{1}{6}[2\bar{1}1]$, respectively. The corresponding disconnection line vectors are $[\bar{1}10]$ and $[\bar{2}1\bar{1}]$.}
  \label{sup_fgr:burgers_circuit_s3}
\end{figure}

Supplementary Figure~\ref{sup_fgr:s3_sd_dd_disconn_lines} shows the \emph{sd} (Supplementary Figure~\ref{sup_fgr:s3_sd_dd_disconn_lines}a) and \emph{dd} (Supplementary Figure~\ref{sup_fgr:s3_sd_dd_disconn_lines}b) of a $\Sigma3$ GB in a nanoparticle. The \emph{sd} line is parallel to $[\bar{2}1\bar{1}]$, and the associated step is most clearly observed in the $[\bar{2}1\bar{1}]$ projection. The Burgers circuits for the \emph{dd} and \emph{sd} of $\Sigma11$ and $\Sigma3$ GBs are shown in Supplementary Figures~\ref{sup_fgr:burgers_circuit_s11} and \ref{sup_fgr:burgers_circuit_s3}, respectively. The corresponding Burgers vector, step height, disconnection line vector, and disconnection type are summarized in Supplementary Table~\ref{sup_tab:disconn_type}.

\begin{figure}[!h]
\centering
  \includegraphics[width=1.0\textwidth]{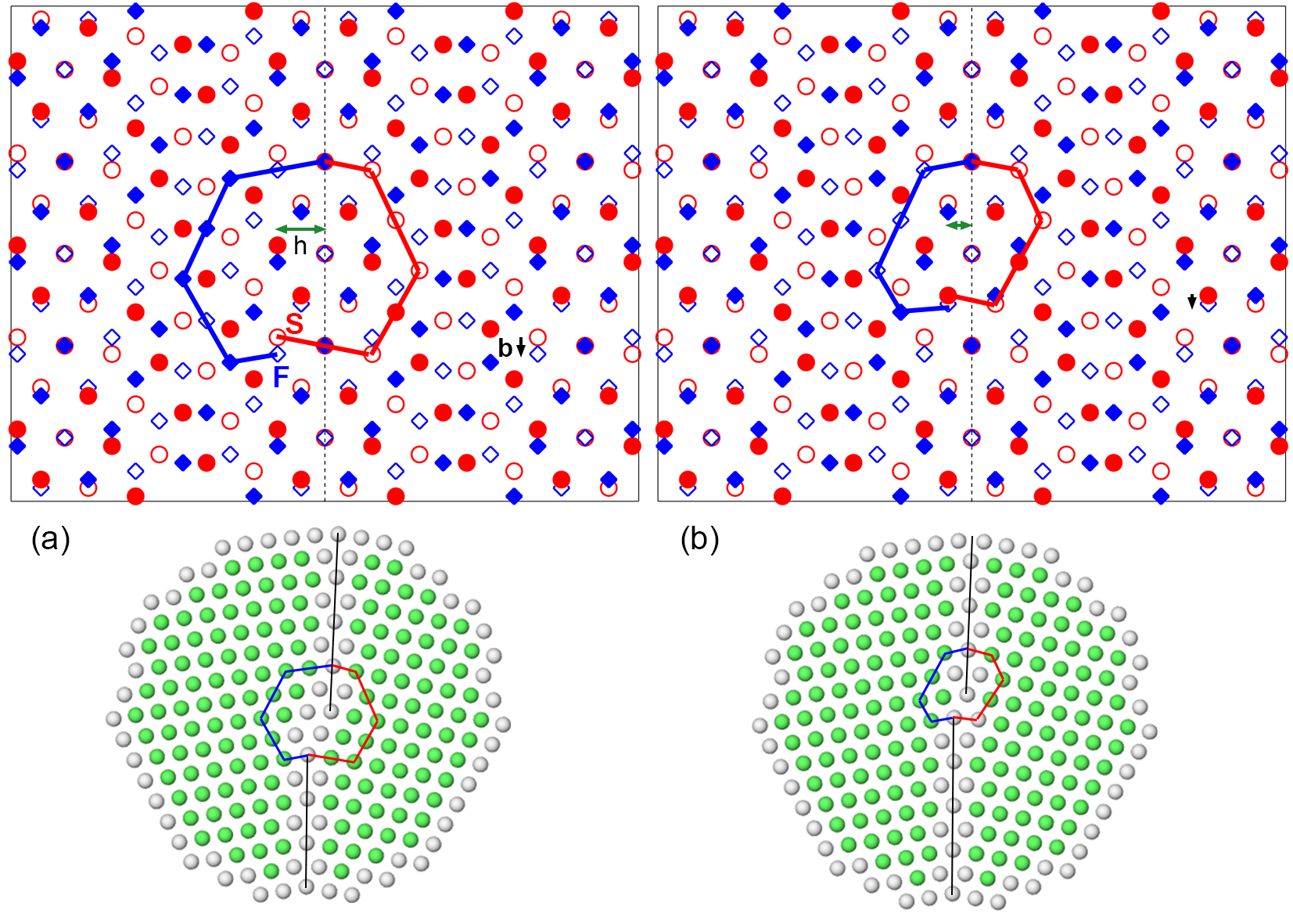}
  \caption{Burgers circuits mapped onto the dichromatic pattern for (a) a $\Sigma11$ double disconnection (\emph{dd}) and (b) a $\Sigma11$ single disconnection (\emph{sd}). In the dichromatic pattern, the right and left crystals are shown by circles and diamonds, respectively. Open and solid symbols refer to heights $z=0$ and $z=+d_{110}$, respectively, where $d_{110}$ is the distance between \{110\} planes. \emph{S} and \emph{F} indicate the start and finish of the Burgers circuit, respectively. $\boldsymbol{b}$ and $h$ are the Burgers vector and step height, respectively. The displacement-shift-complete (DSC) vectors of $\Sigma11$ are $\boldsymbol{D}_x=\frac{1}{11}[113]$, $\boldsymbol{D}_y=\frac{1}{22}[33\bar{2}]$, and $\boldsymbol{D}_z=\frac{1}{4}[\bar{1}10]$. The Burgers vectors of the double disconnection and single disconnection are $\boldsymbol{b}_{dd,\Sigma11}=-\boldsymbol{D}_y=\frac{1}{22}[\bar{3}\bar{3}2]$ and $\boldsymbol{b}_{sd,\Sigma11}=-\frac{\boldsymbol{D}_y}{2}-\boldsymbol{D}_z=\frac{1}{22}[4\bar{7}1]$, respectively. Both $\Sigma11$ disconnections have the same line vector, $[\bar{1}10]$.}
  \label{sup_fgr:burgers_circuit_s11}
\end{figure}

\clearpage
\clearpage
\begin{table}[!h]
\centering
\caption{Burgers vector ($\boldsymbol{b}$), step height ($h$), disconnection line vector ($\boldsymbol{\xi}$), and type of double disconnection (\emph{dd}) and single disconnection (\emph{sd}) of $\Sigma11$ and $\Sigma3$ GBs, corresponding to the Burgers circuits shown in Supplementary Figures~\ref{sup_fgr:burgers_circuit_s11} and \ref{sup_fgr:burgers_circuit_s3}.}
\label{sup_tab:disconn_type}
\def\arraystretch{2.0}
\begin{tabular*}{\textwidth}{@{\extracolsep{\fill}} l c c c c @{}}
\hline
\hline
\textbf{} & $\Sigma11$ (dd) & $\Sigma11$ (sd) & $\Sigma3$ (dd) & $\Sigma3$ (sd) \\
\hline
$\boldsymbol{b}$ & $\frac{1}{22}[\bar{3}\bar{3}2]$ & $\frac{1}{22}[4\bar{7}1]$ & $\frac{1}{6}[\bar{1}\bar{1}\bar{2}]$ & $\frac{1}{6}[2\bar{1}1]$ \\
$h$ & $2d_{113}$ & $d_{113}$ & $2d_{111}$ & $d_{111}$ \\
$\boldsymbol{\xi}$ & $[\bar{1}10]$ & $[\bar{1}10]$ & $[\bar{1}10]$ & $[\bar{2}1\bar{1}]$ \\
type & edge & mixed & edge & screw \\
\hline
\hline
\end{tabular*}
\end{table}

\section{Faulted single disconnections}

The single disconnections of $\Sigma11$ and $\Sigma3$ GBs discussed in the main text are formed through a screw shift, where the regions on the right and left sides undergo a relative displacement equivalent to a stacking change. In contrast, if only one side undergoes the stacking-change displacement, a stacking fault (SF) is required to accommodate the shift. This gives rise to a faulted single disconnection, with the SF emanating from the disconnection. This is shown for $\Sigma11$ and $\Sigma3$ in Supplementary Figure~\ref{sup_fgr:faulted_disconn} (a--c) and (d--f), respectively.

\begin{figure}[!h]
\centering
  \includegraphics[width=1.0\textwidth]{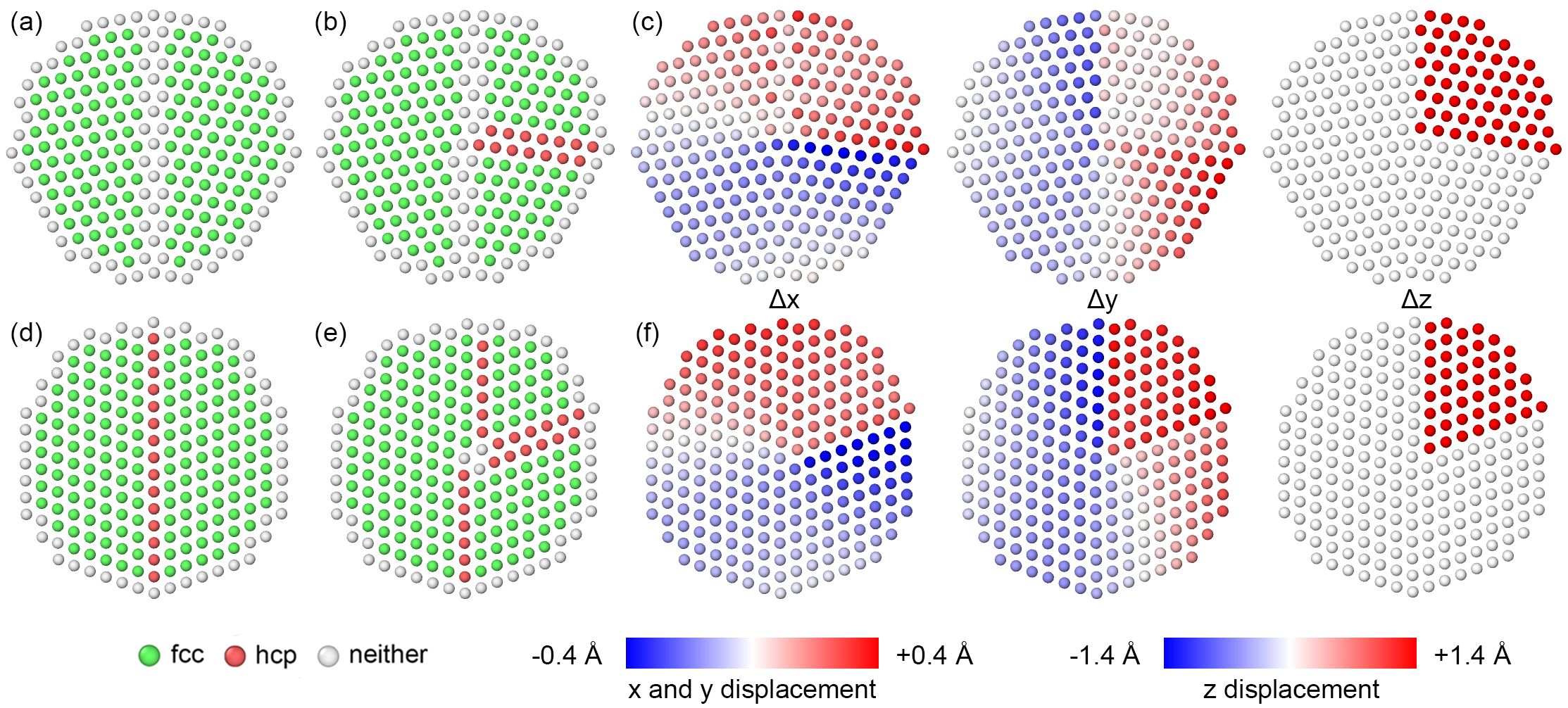}
  \caption{Faulted single disconnections in $\Sigma11$ and $\Sigma3$ GBs. (a) $\Sigma11$ GB in a 2~nm diameter cylinder. (b) Faulted single disconnection in the $\Sigma11$ GB. (c) Corresponding $x$-, $y$-, and $z$-displacement maps. (d) $\Sigma3$ GB in a 2~nm diameter cylinder. (e) Faulted single disconnection in the $\Sigma3$ GB. (f) Corresponding $x$-, $y$-, and $z$-displacement maps.}
  \label{sup_fgr:faulted_disconn}
\end{figure}

\clearpage
\section{Dynamical evolution of kinked column shifts}

\begin{figure}[!h]
\centering
  \includegraphics[width=0.4\textwidth]{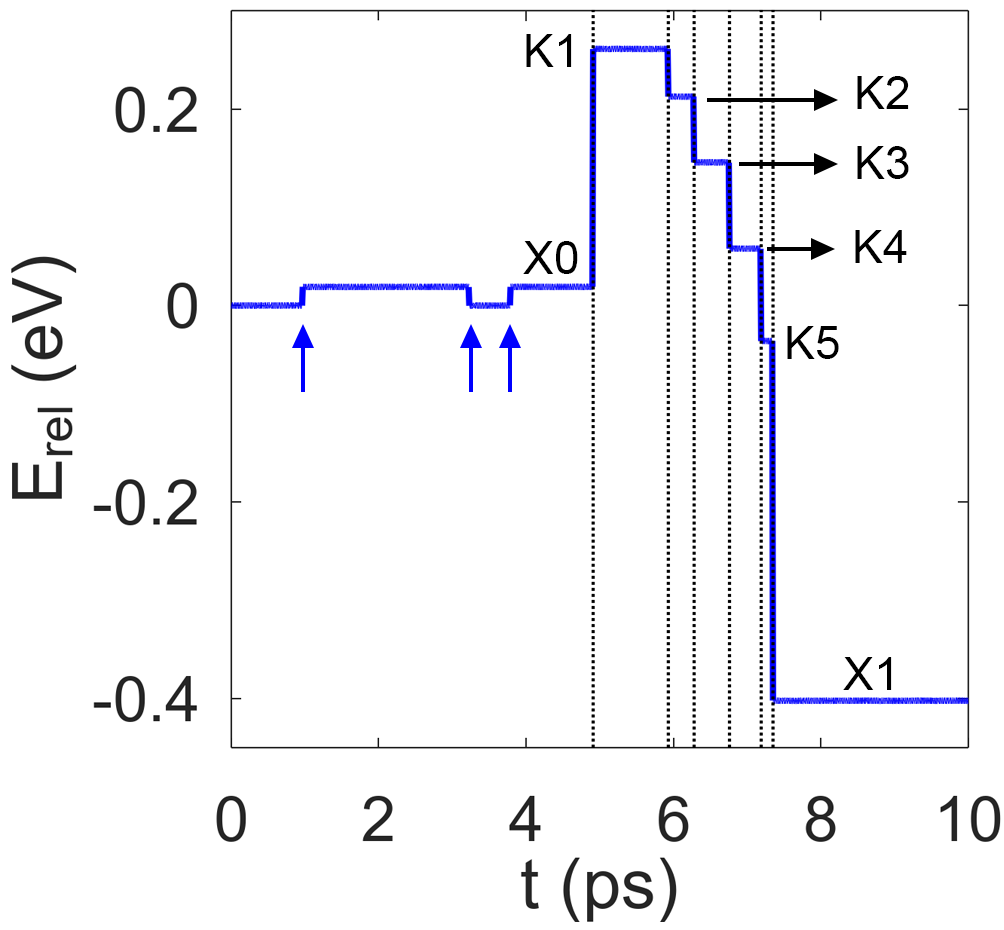}
  \caption{Relaxed energy profile of a 10~ps simulation at 850~K initiated from the configuration at 29.5~ns shown in Figure~\ref{fgr:disconnection_kinks}a. The trajectory undergoes an anti-parallel kinked column shift through configurations X0 $\rightarrow$ K1 $\rightarrow$ K2 $\rightarrow$ K3 $\rightarrow$ K4 $\rightarrow$ K5 $\rightarrow$ X1. The blue arrows indicate surface atom diffusion events preceding X0.}
  \label{sup_fgr:energy_1518_r2}
\end{figure}

Supplementary Figure~\ref{sup_fgr:energy_1518_r2} shows the relaxed (quenched to 0~K) energy profile of a 10~ps trajectory exhibiting a column shift through an anti-parallel kink. After three surface atom diffusion events (indicated by blue arrows), the nanoparticle reaches configuration X0. Independent atomic down shifts then produce the kinked configurations K1 to K5 and finally X1, where the entire column has shifted down. The atoms in the column shift separately, except for the atomic pairs at the two ends of the column, which shift together.

\begin{figure}[!h]
\centering
  \includegraphics[width=0.9\textwidth]{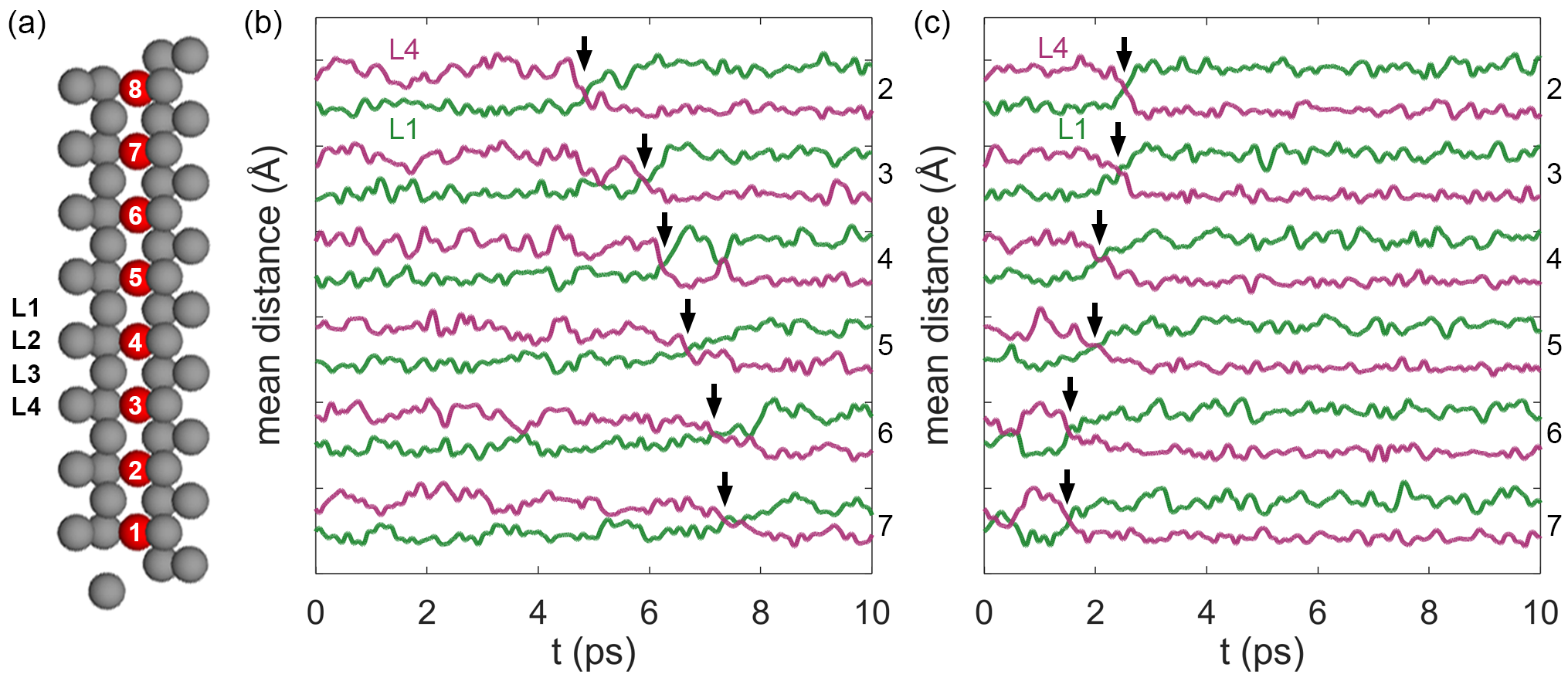}
  \caption{(a) Definition of the neighboring \{110\} layers L1, L2, L3, and L4, illustrated for atom 4 of the shifting column in configuration X0. Mean distances to L1 and L4 for atoms 2 to 7 during (b) an anti-parallel kink and (c) a parallel kink. The crossover of the mean distances to L1 and L4 identifies the time at which each atom undergoes the down shift and is indicated by a black arrow.}
  \label{sup_fgr:displ_mag_L1_L4}
\end{figure}

To resolve the dynamical sequence of these atomic down shifts directly from the finite-temperature trajectories, we measure the mean distance of each shifting atom to reference \{110\} layers on either side of the column. The atoms in the shifting column are labeled 1 to 8 from bottom to top (Supplementary Figure~\ref{sup_fgr:displ_mag_L1_L4}a). For each atom except atoms 1 and 8, four neighboring \{110\} layers, L1, L2, L3, and L4, are defined; this construction is illustrated for atom 4 in Supplementary Figure~\ref{sup_fgr:displ_mag_L1_L4}a. In configuration X0, the shifting atom lies in L2, while after the down shift it lies in L3. Each reference layer consists of three atoms in the neighboring columns (shown in gray). The mean distance to a given layer is calculated as the mean of the distances from the shifting atom to these three atoms. In X0, the mean distance to L1 is smaller than that to L4, while after the down shift this relation is reversed. We therefore identify the time of the atomic shift from the crossover of the mean distances to L1 and L4.

The mean distances to L1 and L4 for atoms 2 to 7 are shown for trajectories exhibiting anti-parallel and parallel kinks in Supplementary Figure~\ref{sup_fgr:displ_mag_L1_L4}b and c, respectively. In the anti-parallel kink, the column shift begins at the bottom, with atomic pair 1--2 shifting first, followed by atoms 3 to 6 individually, and finally atomic pair 7--8. In contrast, for the parallel kink, the shift begins at the top and propagates downward along the column.

\clearpage
\section{NEB atomic pathways}

The evolution of the column shift along the NEB path from X0 to X1 is shown for the three column-shift modes: anti-parallel kink (C\textsubscript{K-}), parallel kink (C\textsubscript{K+}), and full shift (C) in Supplementary Figure~\ref{sup_fgr:neb_images_1518}a, b, and c, respectively. The displacement magnitude ($\Delta s$) of each atom in the shifting column is shown as a function of the displacement magnitude of the center of mass of the column in Supplementary Figure~\ref{sup_fgr:neb_images_1518}d. The NEB pathways shown here are the same as those used for the energy profiles in Figure~\ref{fgr:disconnection_kinks}d. During the full shift, the atoms in the column move together. In the anti-parallel kink pathway, the atoms shift sequentially and largely independently. The parallel kink lies between these two cases, with groups of atoms shifting together as the kink propagates along the column.

\begin{figure}[!h]
\centering
  \includegraphics[width=1.0\textwidth]{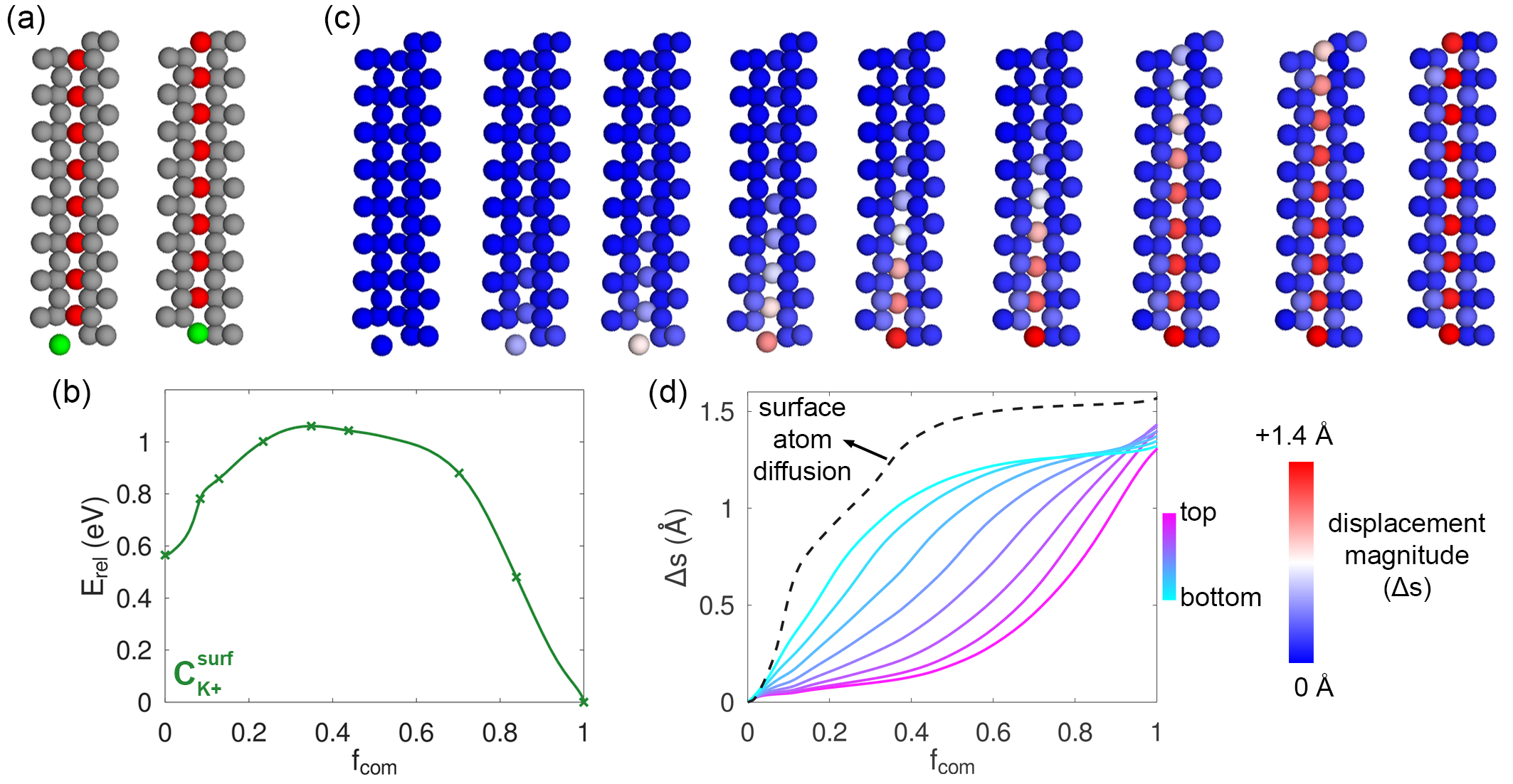}
  \caption{Parallel kink triggered by surface atom diffusion. (a) Initial (X0) and final (X1$^\prime$) configurations; the green atom undergoes surface diffusion into the shifting column. (b) NEB energy profile. (c) Configurations corresponding to the $\times$ markers in panel (b), showing the column down shift. (d) Displacement magnitude of each atom in the shifting column as a function of the displacement magnitude of the center of mass of the original column atoms. The displacement magnitude of the diffusing surface atom is also shown.}
  \label{sup_fgr:surf_diff_par_kink}
\end{figure}

Supplementary Figure~\ref{sup_fgr:surf_diff_par_kink} shows the corresponding pathway for the parallel kink triggered by surface atom diffusion (\( \mathrm{C}_{\mathrm{K}}^{\raisebox{0.50ex}{\scriptsize surf}} \)). As the surface atom enters the shifting column, the NEB energy profile exhibits a small intermediate increase (Supplementary Figure~\ref{sup_fgr:surf_diff_par_kink}b), which is also reflected in the evolution of its displacement magnitude (Supplementary Figure~\ref{sup_fgr:surf_diff_par_kink}d). 

To quantify the differences in the atomic displacement patterns, we define the deviation from simultaneous shift (DSS). For each NEB image $m$,

\begin{equation}
D^{(m)}=\sum_{i<j}\left|u_i^{(m)}-u_j^{(m)}\right|,
\end{equation}

where $u_i^{(m)}$ is the displacement magnitude of atom $i$ in the shifting column. The DSS is then defined as

\begin{equation}
\mathrm{DSS}=\frac{1}{M}\sum_{m=1}^{M}D^{(m)},
\end{equation}

where $M$ is the number of NEB images. A smaller DSS corresponds to a more simultaneous column shift. The DSS values are 10.48~\AA\ for the anti-parallel kink, 3.62~\AA\ for the parallel kink, 0.98~\AA\ for the full shift, and 5.48~\AA\ for the surface diffusion triggered parallel kink. For the latter, only atoms belonging to the original shifting column are included in the calculation.

\begin{figure}[!h]
\centering
  \includegraphics[width=1.0\textwidth]{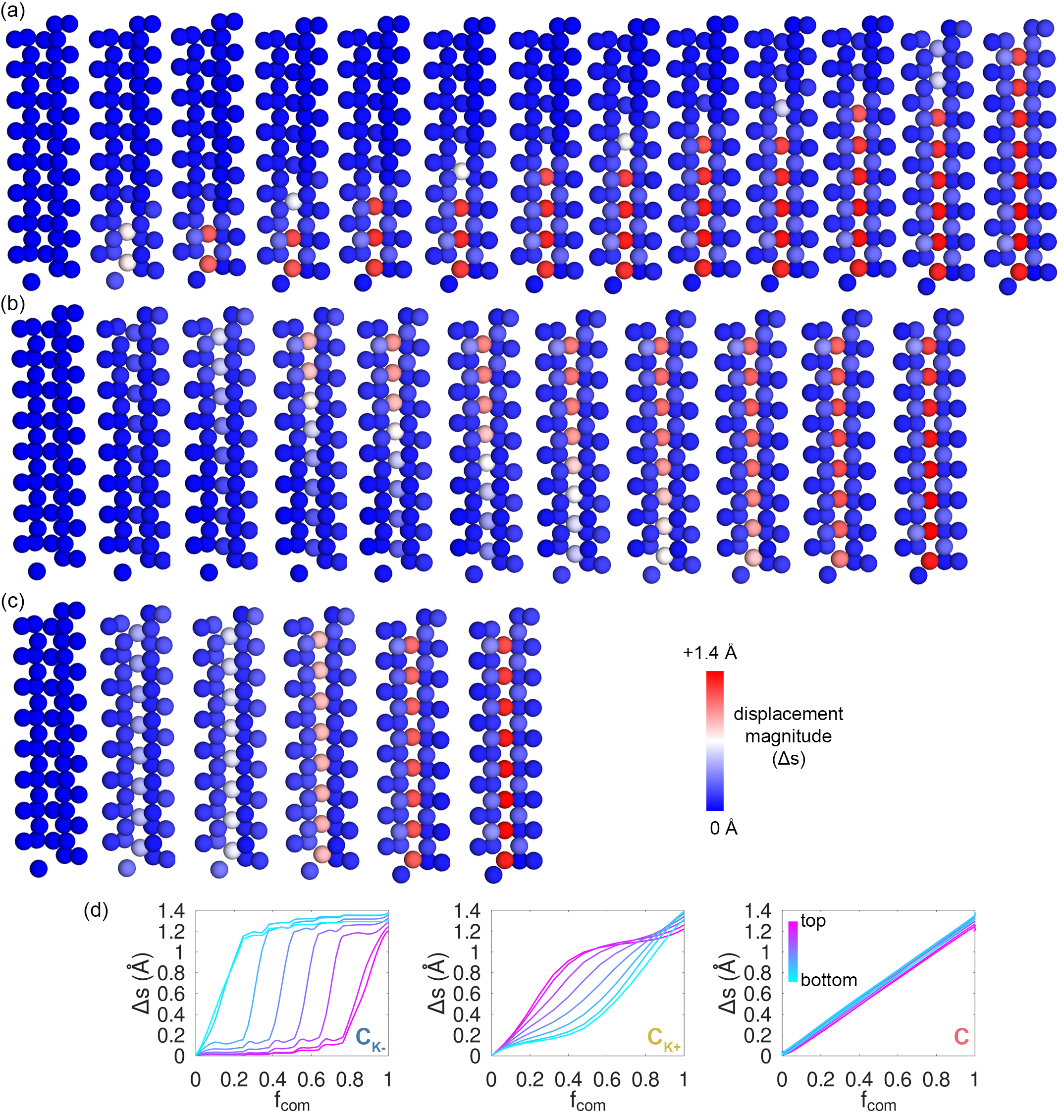}
  \caption{Configurations along the NEB pathways showing the column down shift through (a) an anti-parallel kink, (b) a parallel kink, and (c) a full shift. (d) Displacement magnitude of each atom in the shifting column as a function of the displacement magnitude of the center of mass of the column for the three column shift modes. The atomic displacement curves are colored from magenta for the topmost atom to cyan for the bottommost atom.}
  \label{sup_fgr:neb_images_1518}
\end{figure}

\clearpage
\section{Robustness of the restrained full shift NEB pathway}
\label{sup_sec:kcol_robustness}

Without the additional harmonic restraints, the NEB pathway between X0 and X1 evolves into a kinked column shift instead of retaining the full column shift. Hence, for the full shift, we apply harmonic restraints between neighboring atoms in the shifting column. For each neighboring atomic pair, the restraint energy is

\begin{equation}
E_{\mathrm{rest}}=K_{\mathrm{col}}(r-r_0)^2,
\end{equation}

where $r_0$ is taken as the mean of the corresponding pair distance in the initial and final configurations. The purpose of this restraint is only to prevent the neighboring atoms in the column from entering kinked-like configurations. After the NEB calculation converges, the restraints are removed without further relaxation and the EAM energies of the same configurations are calculated. Therefore, the energy barriers reported here do not include the harmonic restraint energy.

\begin{figure}[!h]
\centering
  \includegraphics[width=0.77\textwidth]{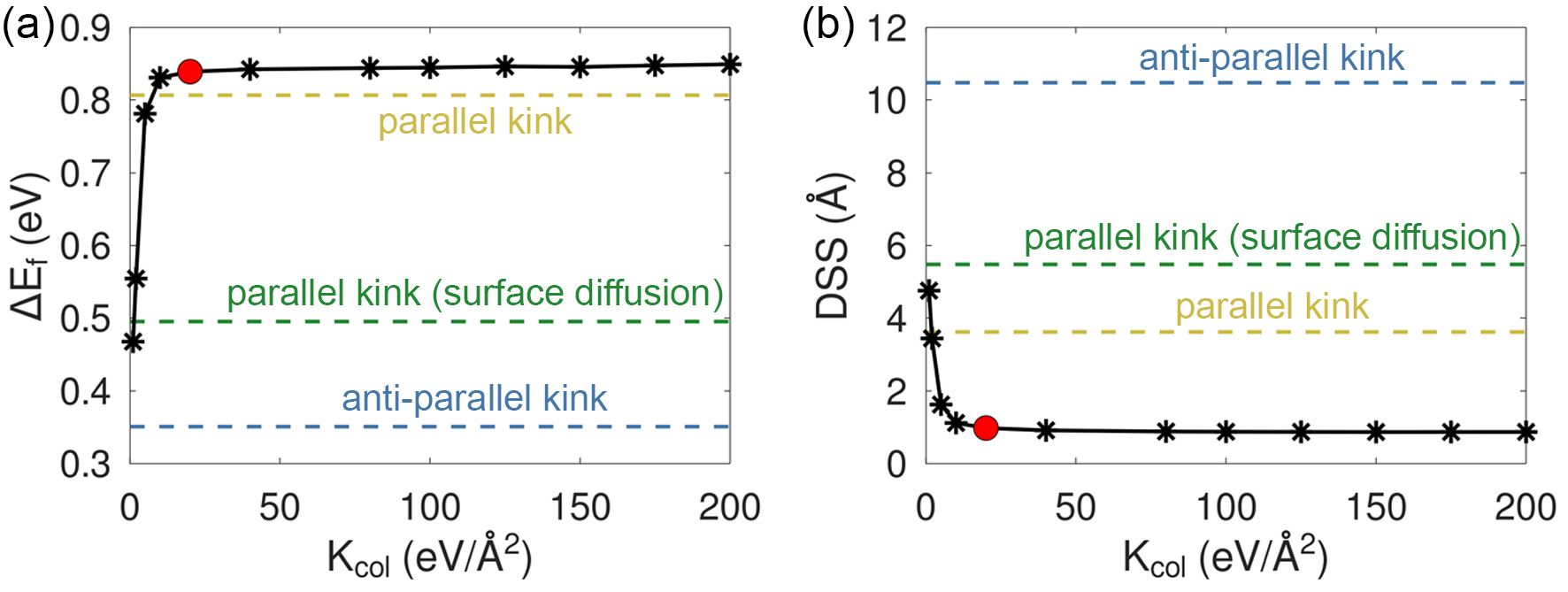}
  \caption{Plots of (a) forward energy barrier ($\Delta{E}_{\mathrm{f}}$) and (b) DSS as a function of $K_{\mathrm{col}}$. Also indicated are the corresponding values for column shifts through anti-parallel kink, parallel kink, and surface diffusion triggered parallel kink. The red marker indicates the values corresponding to $K_{\mathrm{col}}=20$~eV/\AA$^{2}$ in both panels.}
  \label{sup_fgr:kcol_robustness}
\end{figure}

We varied $K_{\mathrm{col}}$ from 1 to 200~eV/\AA$^{2}$ to check how the full-shift pathway and its energy barrier depend on the restraint strength. At low $K_{\mathrm{col}}$, the pathway is not fully concerted and the energy barrier changes strongly with $K_{\mathrm{col}}$. For $K_{\mathrm{col}}\geq20$~eV/\AA$^{2}$, the full column shift is retained and the barrier changes only slightly with increasing $K_{\mathrm{col}}$ (Supplementary Figure~\ref{sup_fgr:kcol_robustness}a).

We also calculate the deviation from simultaneous shift (DSS), defined in the previous section, for each value of $K_{\mathrm{col}}$. The DSS decreases strongly at low $K_{\mathrm{col}}$ and changes only slightly once the full-shift pathway is retained (Supplementary Figure~\ref{sup_fgr:kcol_robustness}b). Based on both the energy barrier and DSS, we use $K_{\mathrm{col}}=20$~eV/\AA$^{2}$ for the full-shift pathway shown in Figure~\ref{fgr:disconnection_kinks}d.

\section{GB processes during 5 ps simulations}

\begin{figure}[!h]
\centering
  \includegraphics[width=1.0\textwidth]{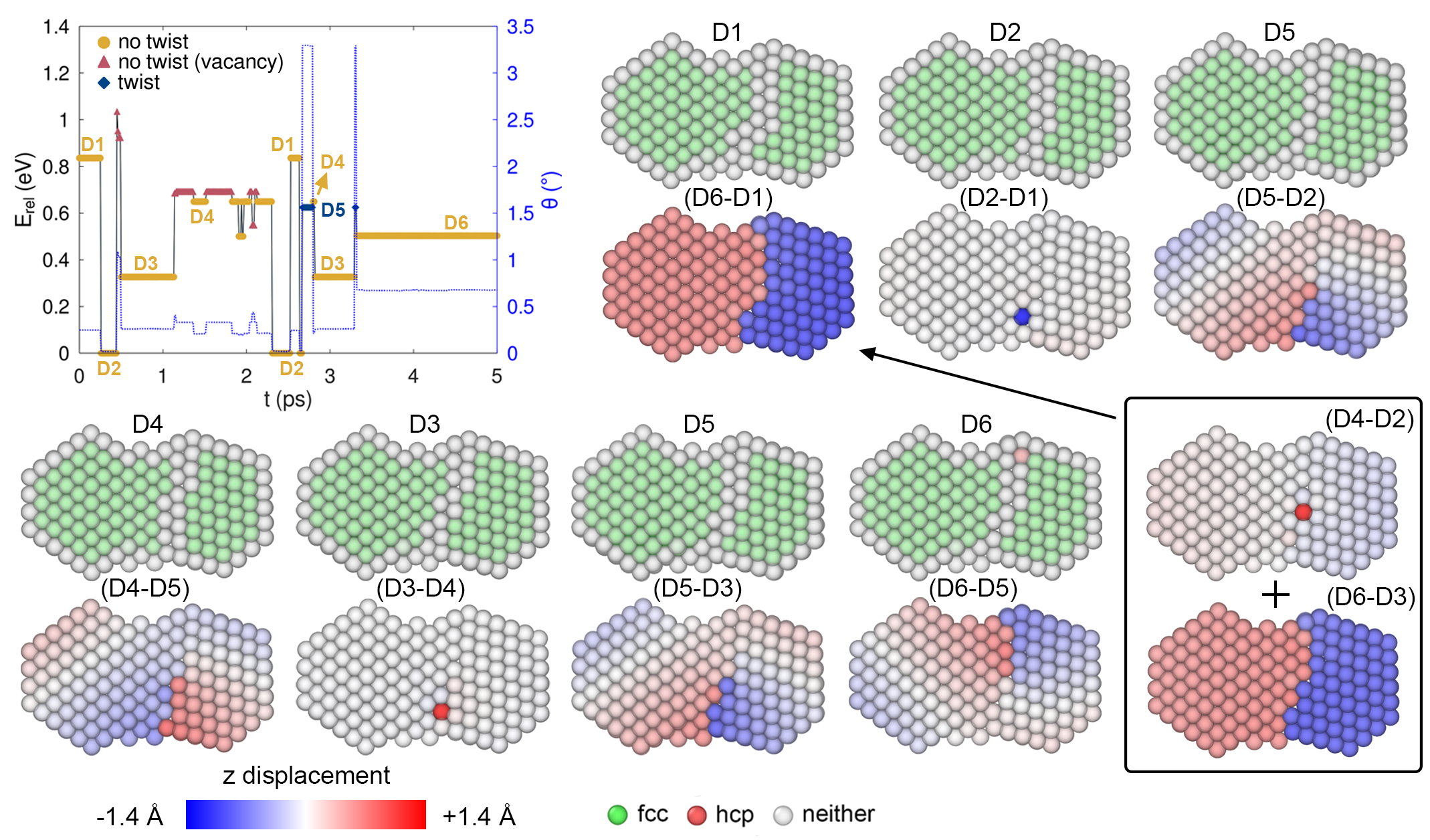}
  \caption{Relaxed energy and twist angle during a 5~ps simulation, together with selected GB configurations and the corresponding $z$-displacement maps. The $z$-displacement map of configuration B with respect to A is denoted as (B-A). This is the 5\textsuperscript{th} simulation out of the 100 simulations.}
  \label{sup_fgr:gb_processes_run5}
\end{figure}

\begin{figure}[!h]
\centering
  \includegraphics[width=1.0\textwidth]{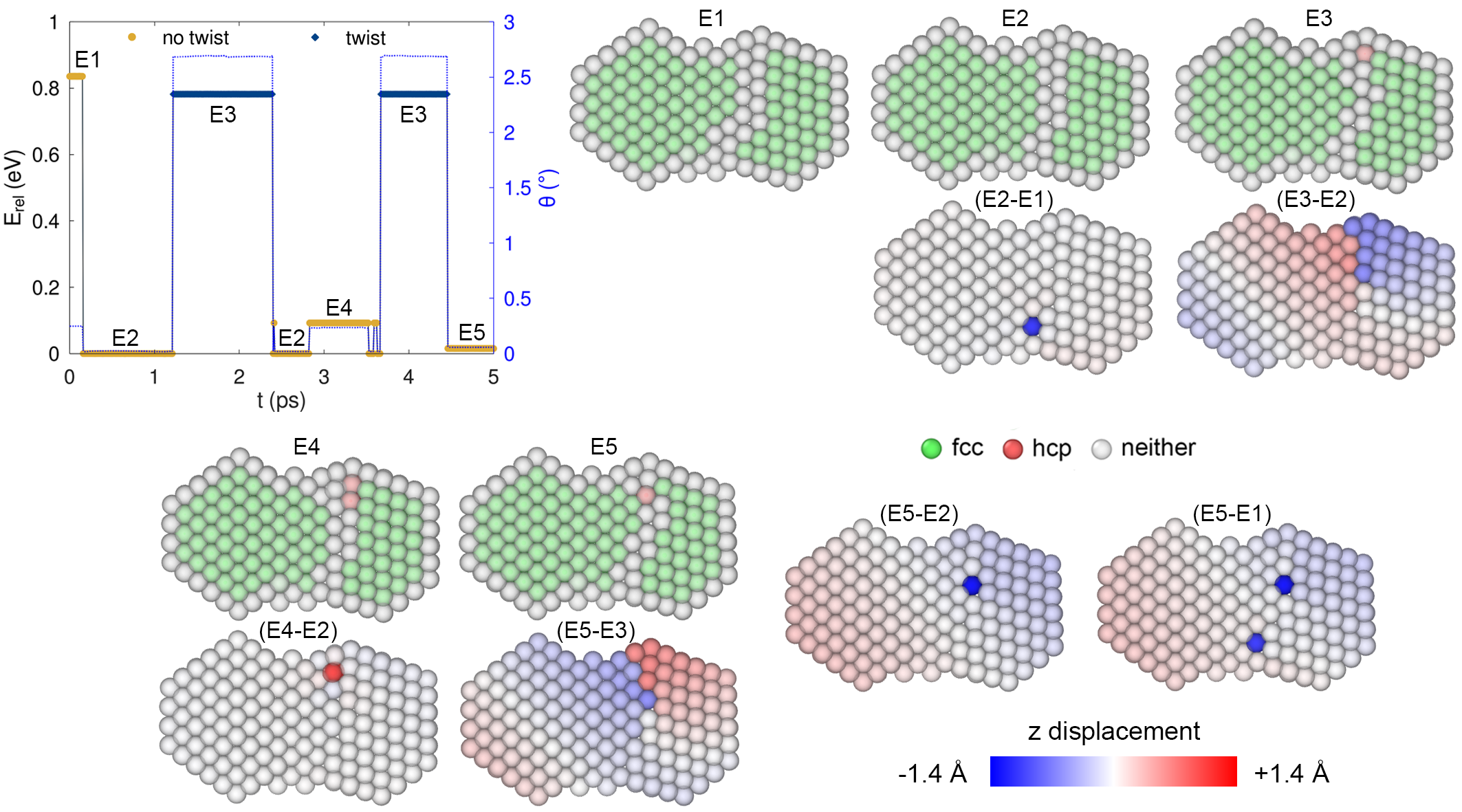}
  \caption{Relaxed energy and twist angle during a 5~ps simulation, together with selected GB configurations and the corresponding $z$-displacement maps. The $z$-displacement map of configuration B with respect to A is denoted as (B-A). This is the 18\textsuperscript{th} simulation out of the 100 simulations.}
  \label{sup_fgr:gb_processes_run18}
\end{figure}

In Figure~\ref{fgr:gb_processes}, we reported the six GB processes observed during the 100 short 5~ps simulations. Here, we provide two examples (Supplementary Figures~\ref{sup_fgr:gb_processes_run5} and \ref{sup_fgr:gb_processes_run18}) to show how these processes occur dynamically during a 5~ps simulation. In each case, we show the changes in the relaxed energy and twist angle ($\theta$), together with configurations showing the various GB processes. The relaxed energy is color-coded according to configurations with no twist (yellow circles), no twist but with stacking-mismatch vacancies, i.e., anti-parallel kinks (red triangles), and configurations with twist that have undergone a partial screw shift (blue diamonds). Configurations with $\theta>1^\circ$ are considered twisted.

In the first example (Supplementary Figure~\ref{sup_fgr:gb_processes_run5}), the initial configuration (D1) undergoes transformations through full (P3) and kinked (P2) column shifts and returns to D1. During this part of the trajectory, we also observe P1, where kinked configurations form but reverse without completing the column shift. From this point onward, the sequence can be described as D1 $\rightarrow$ D2 $\rightarrow$ D5 $\rightarrow$ D4 $\rightarrow$ D3 $\rightarrow$ D5 $\rightarrow$ D6. D1 transforms to D2 through a full column down shift (P3). D2 $\rightarrow$ D5 $\rightarrow$ D4 results in a column up shift in D4 with respect to D2. This corresponds to P4, where two successive partial screw shifts result in a column shift. The twist angle in configuration D5 is slightly larger than $3^\circ$. D4 $\rightarrow$ D3 then occurs through a column up shift (P3). Finally, D3 $\rightarrow$ D5 $\rightarrow$ D6 results in a full screw shift in D6 with respect to D3 (P6).

In the second example (Supplementary Figure~\ref{sup_fgr:gb_processes_run18}), the transformation can be described as E1 (initial configuration) $\rightarrow$ E2 $\rightarrow$ E3 $\rightarrow$ E2 $\rightarrow$ E4 $\rightarrow$ E2 $\rightarrow$ E3 $\rightarrow$ E5. E1 transforms to E2 through a full column down shift (P3). E2 $\rightarrow$ E3 $\rightarrow$ E2 represents a screw-shift reversal (P5), where E3 is a twisted configuration with $\theta>2^\circ$. E2 $\rightarrow$ E4 occurs through a full column up shift (P3), but this column shift then reverses to recover E2. Finally, E2 $\rightarrow$ E3 $\rightarrow$ E5 represents a full column shift through two successive partial screw shifts (P4).

Across these two examples, all six GB processes defined in Figure~\ref{fgr:gb_processes} are observed.

\clearpage
\section{Energy barriers of competing GB pathways}

For the GB processes P2, P4, and P6 described in Figure~\ref{fgr:gb_processes}, we also calculated the barriers for competing pathways that were not observed during the simulations. For P2 (X0 $\rightarrow$ X1, column up shift through anti-parallel kink), we considered an up shift through parallel kink and down shifts through anti-parallel and parallel kinks. For P4 (X0 $\rightarrow$ X1, column down shift through successive partial screw shifts), we considered the down shift through anti-parallel and parallel kinks. These barriers are listed in Supplementary Table~\ref{sup_tab:barrier_gb_process}. For the observed P4 and P6 processes, we also report the barriers for the individual partial screw shifts. For P6, we additionally attempted to construct a direct X0 $\rightarrow$ XF pathway, but the NEB pathway consistently evolved through the intermediate XS configuration.

\begin{table}[!h]
\centering
\caption{Forward ($\Delta$E\textsubscript{f}) and backward ($\Delta$E\textsubscript{b}) energy barriers along with the energy difference between the final and the initial configurations ($\Delta$E) for some of the GB processes reported in Figure~\ref{fgr:gb_processes}. (*) denotes competing pathways that were manually constructed. K-, K+, and S represent anti-parallel kink, parallel kink, and screw shift mechanisms, respectively. Up and down denote the column shift direction where applicable.}
\def\arraystretch{2.0}
\begin{tabular*}{\textwidth}{@{\extracolsep{\fill}} l c c c c @{}}
\hline
\hline
Process &  Mechanism & $\Delta$E\textsubscript{f} (eV) & $\Delta$E\textsubscript{b} (eV) & $\Delta$E (eV) \\
\hline
P2 (X0 $\rightarrow$ X1, up) & K- & 0.371 & 0.880 & -0.509 \\
(*) (X0 $\rightarrow$ X1, up) & K+ & 0.399 & 0.908 & -0.509 \\
(*) (X0 $\rightarrow$ X1', down) & K- & 0.500 & 1.449 & -0.949 \\
(*) (X0 $\rightarrow$ X1', down) & K+ & 0.613 & 1.561 & -0.949 \\
\hline
P4 (X0 $\rightarrow$ X1, down) & S & 0.277 & 0.926 & -0.649 \\
(X0 $\rightarrow$ XS) & S & 0.276 & 0.301 & -0.025 \\
(XS $\rightarrow$ X1) & S & 0.301 & 0.926 & -0.624 \\
(*) (X0 $\rightarrow$ X1, down) & K- & 0.275 & 0.925 & -0.649 \\
(*) (X0 $\rightarrow$ X1, down) & K+ & 0.283 & 0.932 & -0.649 \\
\hline
P6 (X0 $\rightarrow$ XF) & S & 0.601 & 0.425 & 0.176 \\
(X0 $\rightarrow$ XS) & S & 0.601 & 0.303 & 0.298 \\
(XS $\rightarrow$ XF) & S & 0.255 & 0.377 & -0.122 \\
\hline
\hline
\end{tabular*}
\label{sup_tab:barrier_gb_process}
\end{table}

\clearpage
\section{Dynamical atomic displacements in $\Sigma3$ GB}

In the Supplementary Figure \ref{sup_fgr:atomic_displacements_s3}, we demonstrate the dynamical C and S displacements observed during the elimination of a \emph{dd} in a $\Sigma3$ GB, resulting in a symmetric $\Sigma3$ GB. In Supplementary Figure \ref{sup_fgr:atomic_displacements_s3}a, the \emph{dd} present at 130 ps is eliminated through a series of column shifts. These are all down shifts, as indicated by the $z$-displacement maps. Some of these column shifts reverse during the trajectory, but overall, a series of columns belonging to a \{111\} plane undergo down shifts, resulting in the migration of the \emph{dd} out of the particle. Some of these column shifts occur through anti-parallel kinks, as shown in Supplementary Figure \ref{sup_fgr:atomic_displacements_s3}b.

We also observed a few instances where an \emph{sd} formed from a symmetric $\Sigma3$ GB and then reversed. One such instance, between configurations at 598 ps and 599 ps, is shown in the $\langle110\rangle$ (Supplementary Figure \ref{sup_fgr:atomic_displacements_s3}c) and $\langle112\rangle$ (Supplementary Figure \ref{sup_fgr:atomic_displacements_s3}d) projections. The S displacement along the $\langle112\rangle$ direction is clearly visible in the $z$-displacement map in Supplementary Figure \ref{sup_fgr:atomic_displacements_s3}d.

\begin{figure}[!h]
\centering
  \includegraphics[width=0.77\textwidth]{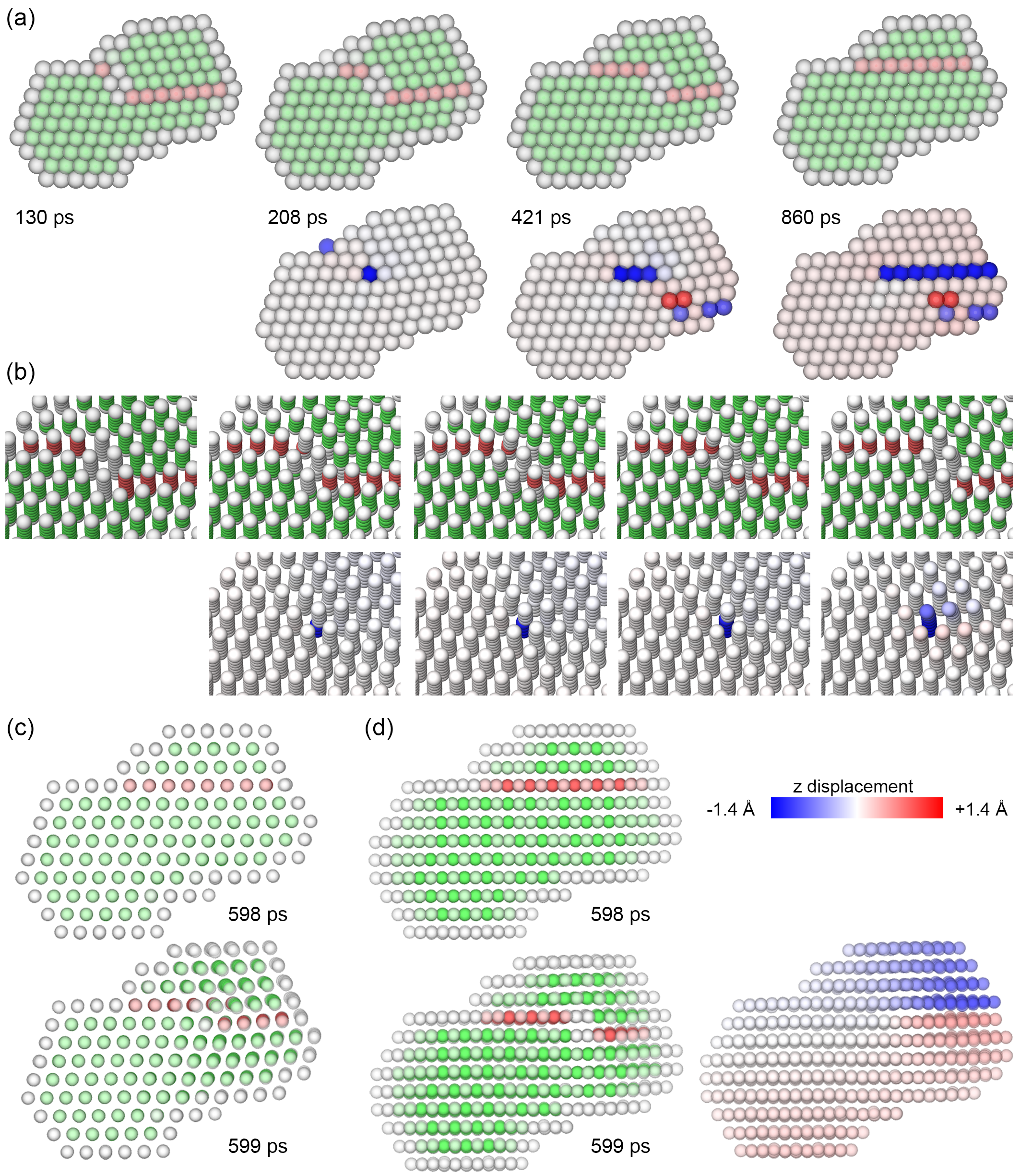}
  \caption{(a) Representative snapshots showing the migration of a \emph{dd} out of the particle through a series of column down shifts. The $z$-displacement maps are calculated with respect to the configuration at 130 ps. The other $z$-displacements, apart from those of the series of shifting columns, correspond to surface atom diffusion events. (b) Migration of the \emph{dd} through a column down shift showing anti-parallel kink configurations. The structures are visualized along a direction slightly tilted away from the $\langle110\rangle$ tilt axis. Transformation of a symmetric $\Sigma3$ GB (598 ps) to a $\Sigma3$ GB with a \emph{sd} (599 ps), visualized in (c) $\langle110\rangle$ and (d) $\langle112\rangle$ projections.}
  \label{sup_fgr:atomic_displacements_s3}
\end{figure}

\clearpage
\section{Geometric transformation of symmetric $\Sigma11$ to symmetric $\Sigma3$}
In Supplementary Figure \ref{sup_fgr:cyl_s11_to_s3}, we demonstrate the transformation of a symmetric $\Sigma11$ GB to a symmetric $\Sigma3$ GB in a 2~nm cylinder. During the transformation, we also calculate the rotation field in the 2D projection along the tilt axis (same as the cylinder axis), following the procedure reported in Ref.~\cite{settem2024_GBJ_disclinations}. The 2D projection of each atomic column is obtained by taking the mean \((x,y)\) position of all the atoms in that column. The rotation field is defined only for projected columns in which all atoms initially have fcc coordination. Thus, the columns belonging to the GB and surface, shown as smaller atoms, are assigned zero rotation. Each of the two grains in the initial symmetric $\Sigma11$ GB has its own reference for calculating the local rotation field (see Ref.~\cite{settem2024_GBJ_disclinations} for details). As the GB evolves, the connected regions originating from each initial grain retain the same reference.

Supplementary Figure \ref{sup_fgr:cyl_s11_to_s3}a shows the symmetric $\Sigma11$ GB and the corresponding rotation field, which is approximately \(0^\circ\) in both grains. Through a suitable choice of stacking change column shifts, the GB transforms through configurations b1, b2, and b3 (Supplementary Figure \ref{sup_fgr:cyl_s11_to_s3}b). In b1, the upper and lower parts of the initial $\Sigma11$ GB migrate laterally towards the left and right, respectively, and in the process generate a $\Sigma3$ segment connecting the two $\Sigma11$ segments. This can also be viewed as two $\Sigma3$--$\Sigma11$ double junction disclinations connected by a common $\Sigma3$ segment. In b2, the structure consists of a $\Sigma3$ GB with a double disconnection, which ultimately transforms to a symmetric $\Sigma3$ GB in b3.

This transformation requires significant rotational accommodation within the particle. The obtuse angle between the \{100\} planes (indicated by black lines) changes from \(129.52^\circ\) in the initial symmetric $\Sigma11$ configuration to \(109.47^\circ\) in the final symmetric $\Sigma3$ configuration, corresponding to a reduction of about \(20^\circ\). The rotation fields of b1 and b3 are shown in Supplementary Figure \ref{sup_fgr:cyl_s11_to_s3}c. Already in b1, the regions near the $\Sigma3$--$\Sigma11$ junctions show noticeable local rotation. In b3, the upper and lower regions show rotations of about \(10^\circ\) in opposite directions, which together account for the required reduction in the obtuse angle during the transformation.

\begin{figure}[!h]
\centering
  \includegraphics[width=1.0\textwidth]{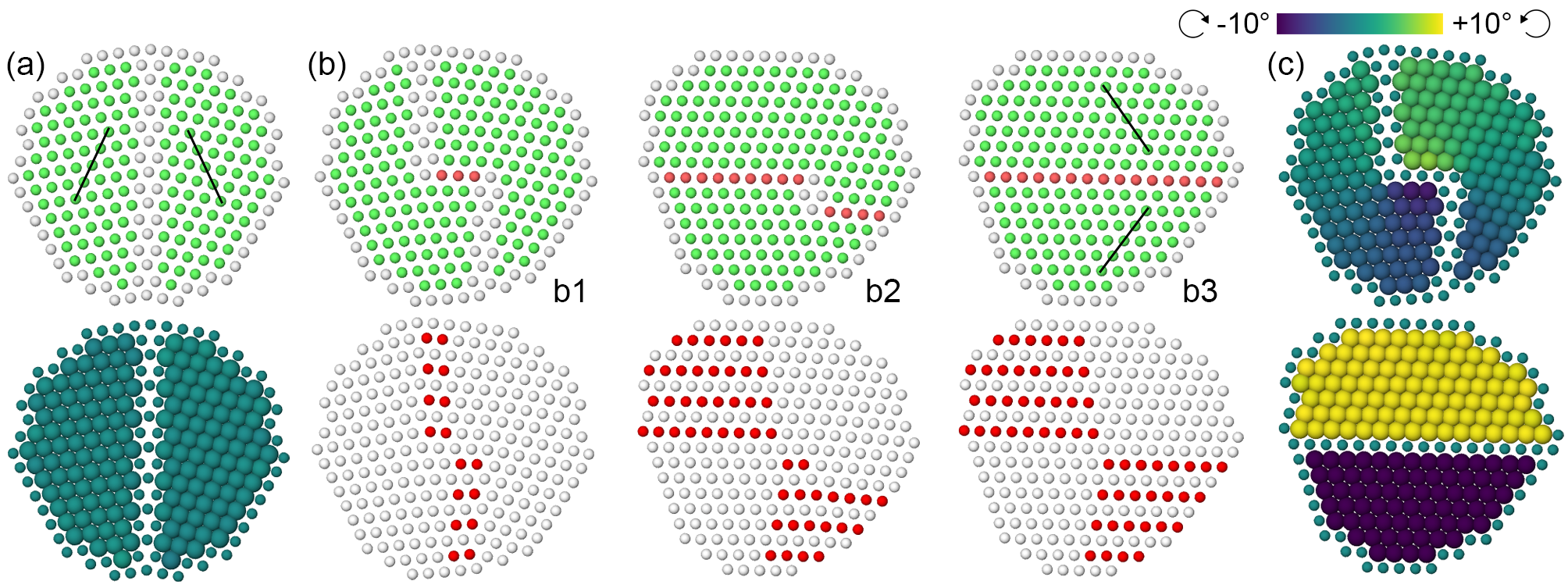}
  \caption{(a) to (c) depicts the transformation from a symmetric $\Sigma$11 to symmetric $\Sigma$3 grain boundary in a 2 nm diameter cylinder through stacking-change column shifts alone (amrked in red). (a) Symmetric $\Sigma$11 grain boundary along with its rotation field. (b) grain boundary structure and the column shifts undergoing stacking change (in red) in configurations b1, b2, and b3. (c) The rotation field in configurations b1 and b3. The black lines in panels (a) and (b) indicate \{100\} planes.}  \label{sup_fgr:cyl_s11_to_s3}
\end{figure}

\clearpage
\section{Structural changes during the $\Sigma11 \rightarrow \Sigma3$ transformation}

\begin{figure}[!h]
\centering
  \includegraphics[width=0.9\textwidth]{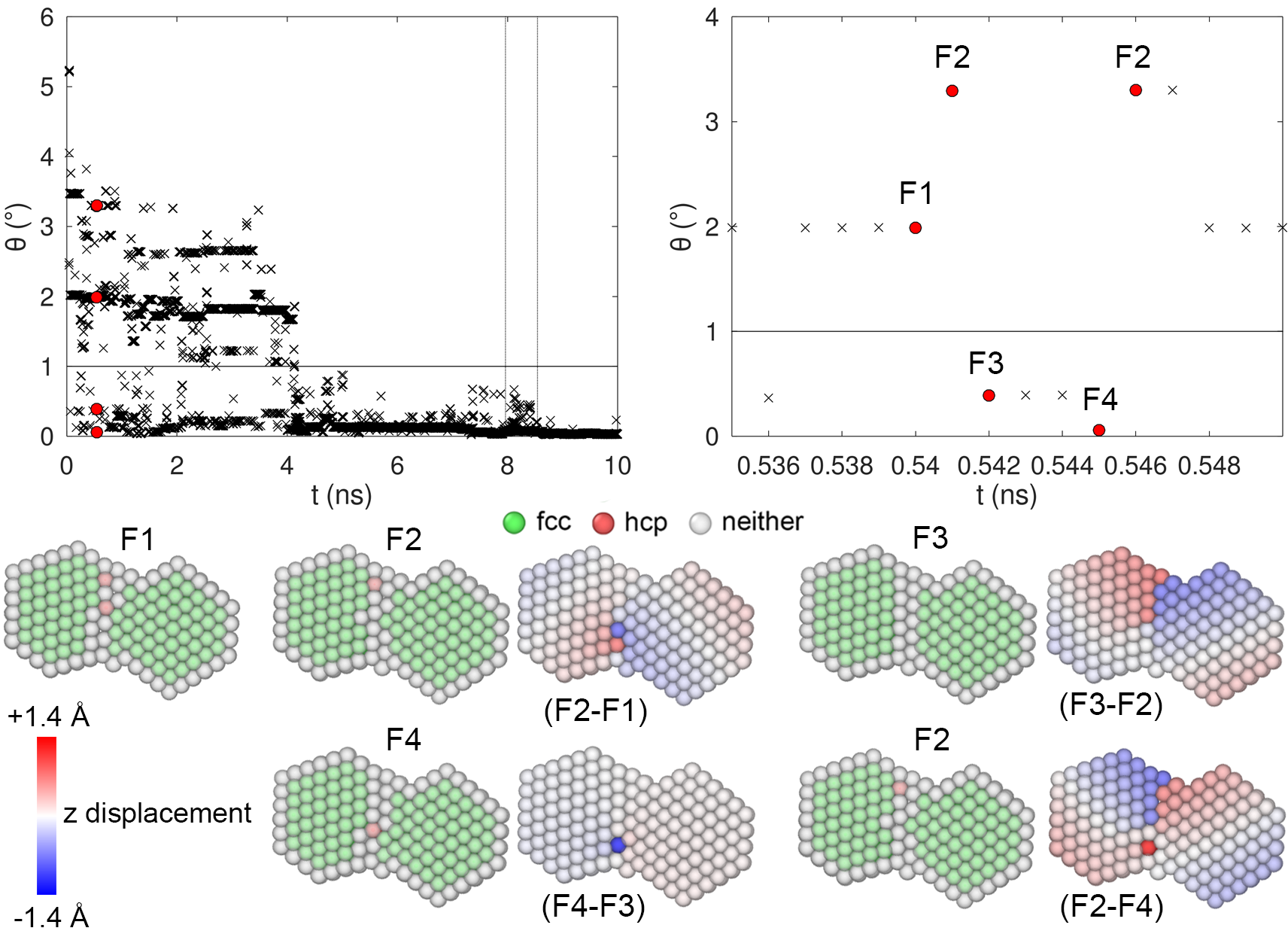}
  \caption{Zoom of the local sequence corresponding to the red markers in the twist angle $\theta$, together with the associated structural changes. The $z$-displacement map of configuration B with respect to A is denoted as (B-A).}
  \label{sup_fgr:zoom_rand_r13}
\end{figure}

In Figure~\ref{fgr:s11_to_s3}c, we marked a local sequence involving partial screw shifts with red markers. Supplementary Figure~\ref{sup_fgr:zoom_rand_r13} shows a zoomed time window corresponding to this sequence, with the structural changes described as F1 $\rightarrow$ F2 $\rightarrow$ F3 $\rightarrow$ F4 $\rightarrow$ F2. F1 is already a twisted configuration with $\theta\sim2^\circ$ and transforms to another twisted configuration F2 ($\theta>3^\circ$) through a highly localized screw shift involving only two atomic columns, where one column shifts up and the other shifts down. A further partial screw shift in the upper region results in F3, which is no longer a twisted configuration ($\theta<1^\circ$). F3 $\rightarrow$ F4 occurs through a column down shift. Reversal of this column shift, followed by reversal of the partial screw shift involved in F2 $\rightarrow$ F3, results in F2 again.

The first significant change in $\theta_{mis}$ observed in Figure~\ref{fgr:s11_to_s3}b is detailed in Supplementary Figure~\ref{sup_fgr:mis_step_rand_r13}. A zoom of this time region for the acute angle is shown in Supplementary Figure~\ref{sup_fgr:mis_step_rand_r13}a, where three locations labeled 1, 2, and 3 are marked. The corresponding structures are shown in Supplementary Figure~\ref{sup_fgr:mis_step_rand_r13}b. The transformation $1\rightarrow2$ occurs through displacement of atoms belonging to a surface \{111\} facet (see the displacement map of configuration 2 with respect to configuration 1). The atoms in the facet do not necessarily move simultaneously, but can rearrange along different $\langle112\rangle$ directions. This results in a subsurface $\Sigma3$ GB. A subsequent column down shift results in configuration 3. During the transformation from $1\rightarrow3$, $\theta_{mis}$ increases by $\sim3^\circ$.

The GB structural details of the configurations at 7963~ps and 8014~ps shown in Figure~\ref{fgr:s11_to_s3}d are given in Supplementary Figure~\ref{sup_fgr:junction_defects}a and b, respectively.

\begin{figure}[!h]
\centering
  \includegraphics[width=1.0\textwidth]{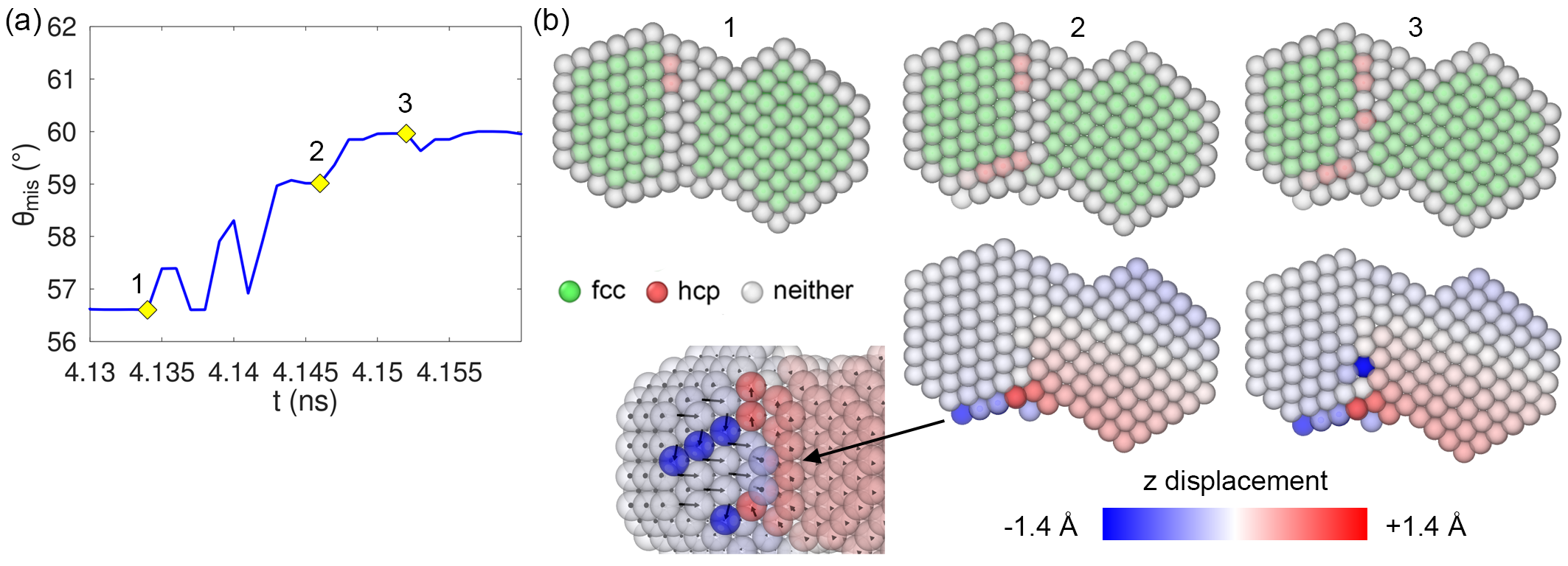}
  \caption{(a) Zoom of the time window where the first significant change in $\theta_{mis}$ is observed. (b) Configurations 1, 2, and 3 (marked in panel a), showing the structural changes during this time window.}
  \label{sup_fgr:mis_step_rand_r13}
\end{figure}

\begin{figure}[!h]
\centering
  \includegraphics[width=0.65\textwidth]{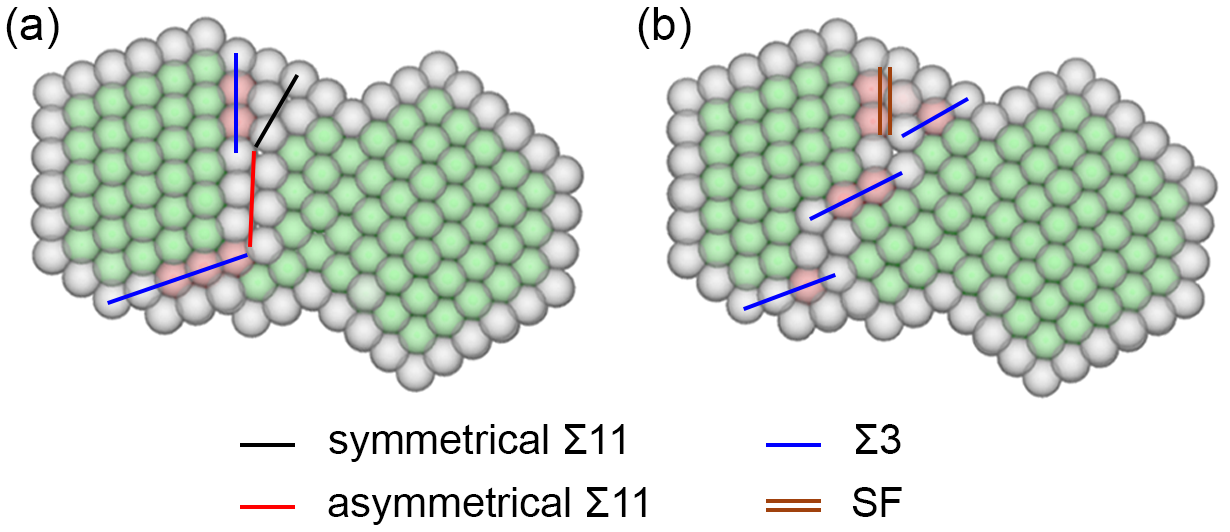}
  \caption{GB configurations at (a) 7963~ps and (b) 8014~ps, highlighting the different GB segments and stacking fault.}
  \label{sup_fgr:junction_defects}
\end{figure}

\clearpage
\section{Additional examples of $\Sigma3$, $\Sigma11$, and junction evolution}

Here, we present three additional examples of GB evolution involving $\Sigma3$, $\Sigma11$, and their junctions.

In Supplementary Figure~\ref{sup_fgr:DJ_to_s3}, we show a transformation from a $\Sigma3-\Sigma11$ double junction disclination to a $\Sigma3$ GB. The obtuse misorientation angle, $\theta_{mis}$, decreases continuously and eventually reaches the value corresponding to $\Sigma3$ (Supplementary Figure~\ref{sup_fgr:DJ_to_s3}a). Representative configurations show the intermediate GB structures during this transformation (Supplementary Figure~\ref{sup_fgr:DJ_to_s3}b). In this case, we also observe that the two lattice regions translate in opposite directions along the tilt axis (Supplementary Figure~\ref{sup_fgr:DJ_to_s3}c). To quantify this translation, we select one reference atomic column in each lattice region and measure the $z$ coordinates of their centers of mass. These are denoted by $z_r$ and $z_l$ for the right and left reference columns, respectively, and the relative translation is measured as $z_r-z_l$, with its initial value set to zero. The evolution of $z_r-z_l$ for the transformations shown in Figure~\ref{fgr:s11_to_s3} and Supplementary Figure~\ref{sup_fgr:DJ_to_s3} is shown in Supplementary Figure~\ref{sup_fgr:zdiff_r13_r72}a and b, respectively. For the transformation in Figure~\ref{fgr:s11_to_s3}, there is no net relative translation at the end of the transformation. In contrast, for the transformation in Supplementary Figure~\ref{sup_fgr:DJ_to_s3}, the right lattice region shifts downward with respect to the left region, with the relative shifts occurring approximately in multiples of the \{110\} interplanar distance, $d_{110}$.

In Supplementary Figure~\ref{sup_fgr:DJ_to_s3_s11_migration_r49}, we show the transformation of a faceted $\Sigma11$ GB, consisting of symmetric and asymmetric segments, to a $\Sigma3$ GB. The initial faceted $\Sigma11$ develops a $\Sigma3-\Sigma11$ double junction disclination. The $\Sigma11$ segment of this junction migrates towards the particle periphery and then transforms to a $\Sigma3$ GB containing a \emph{dd}. Finally, the \emph{dd} migrates out of the particle, leaving behind a $\Sigma3$ GB.

In Supplementary Figure~\ref{sup_fgr:DJ_hits_s3_to_TJ_r23}, we show how a $\Sigma3-\Sigma11$ double junction disclination impinges on a $\Sigma3$ GB, resulting in a $\Sigma3-\Sigma3-\Sigma11$ triple junction disclination. This simulation differs from those described in the Methods section. Here, coalescence was carried out between two randomly oriented 405-atom nanoparticles, each containing a CTB. After the transient rearrangement following attachment, a $\Sigma3-\Sigma11$ double junction and an isolated $\Sigma3$ GB are present. A stacking fault initially emanates from the $\Sigma3$ segment of the double junction and is subsequently eliminated through a screw shift. The $\Sigma11$ segment of the double junction then migrates towards the isolated $\Sigma3$ GB, resulting in migration of the double junction disclination. Eventually, the double junction impinges on the isolated $\Sigma3$ GB and forms a $\Sigma3-\Sigma3-\Sigma11$ triple junction disclination.

In all three examples, we verified that the GB evolution occurs through the C and S displacements.

\begin{figure}[!h]
\centering
  \includegraphics[width=1.0\textwidth]{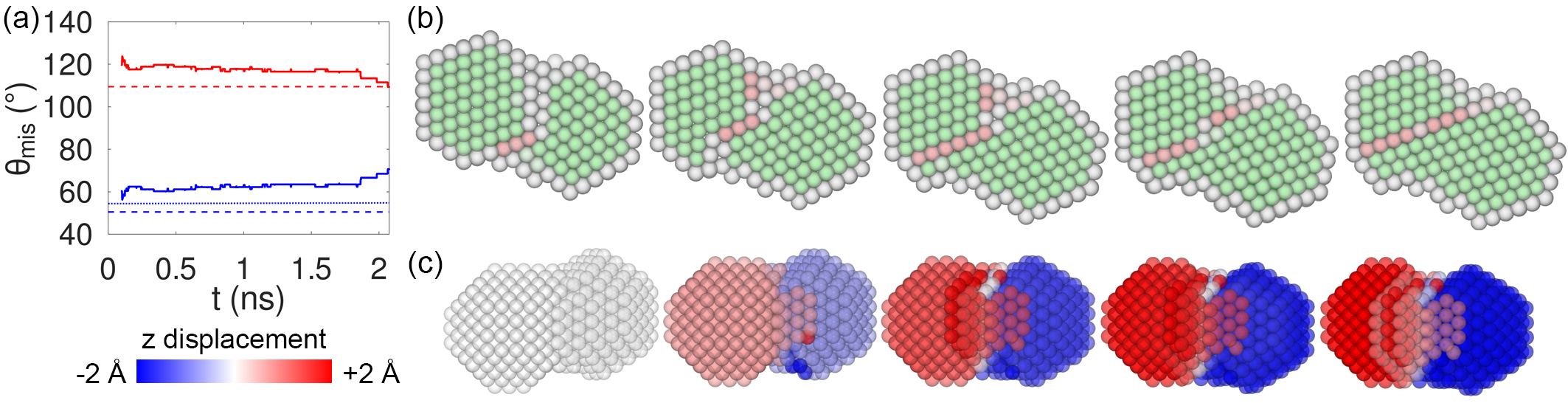}
  \caption{Transformation of a $\Sigma3-\Sigma11$ double junction disclination to a $\Sigma3$ GB. (a) Evolution of the misorientation angle $\theta_{mis}$. (b) Representative structural changes during the transformation. (c) Side views showing the relative translation of the two lattice regions along the tilt axis.}
  \label{sup_fgr:DJ_to_s3}
\end{figure}

\begin{figure}[!h]
\centering
  \includegraphics[width=0.82\textwidth]{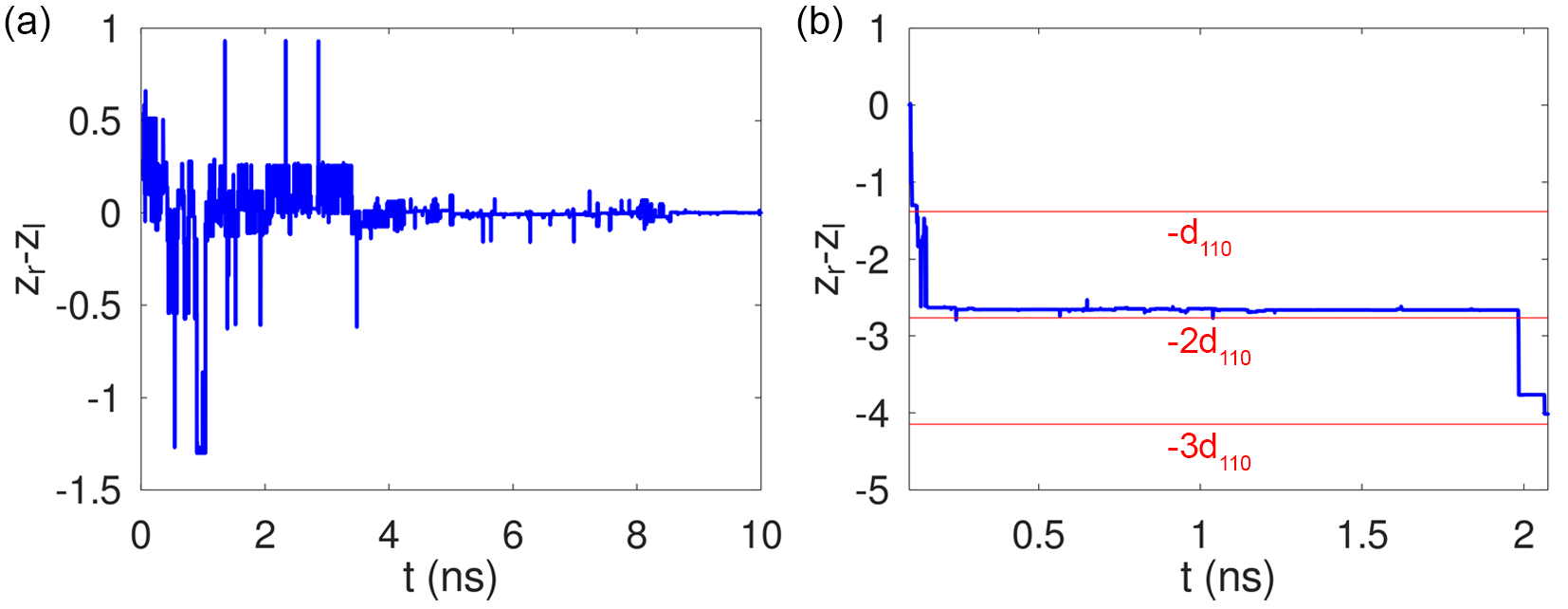}
  \caption{Relative translation, $z_r-z_l$, between reference atomic columns in the right and left lattice regions during the transformations shown in (a) Figure~\ref{fgr:s11_to_s3} and (b) Supplementary Figure~\ref{sup_fgr:DJ_to_s3}. The initial value of $z_r-z_l$ is set to zero.}
  \label{sup_fgr:zdiff_r13_r72}
\end{figure}

\begin{figure}[!h]
\centering
  \includegraphics[width=1.0\textwidth]{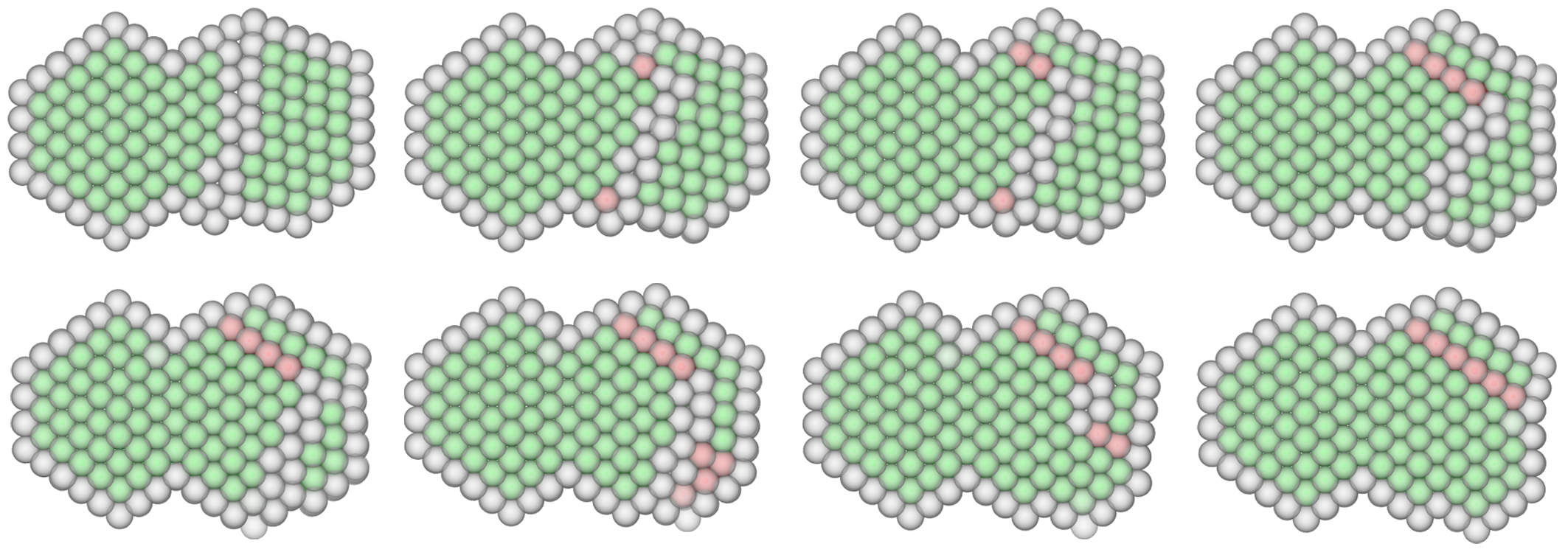}
  \caption{Structural changes during the transformation of a faceted $\Sigma11$ GB to a $\Sigma3$ GB through the formation and migration of a $\Sigma3-\Sigma11$ double junction disclination.}
  \label{sup_fgr:DJ_to_s3_s11_migration_r49}
\end{figure}

\begin{figure}[!t]
\centering
  \includegraphics[width=1.0\textwidth]{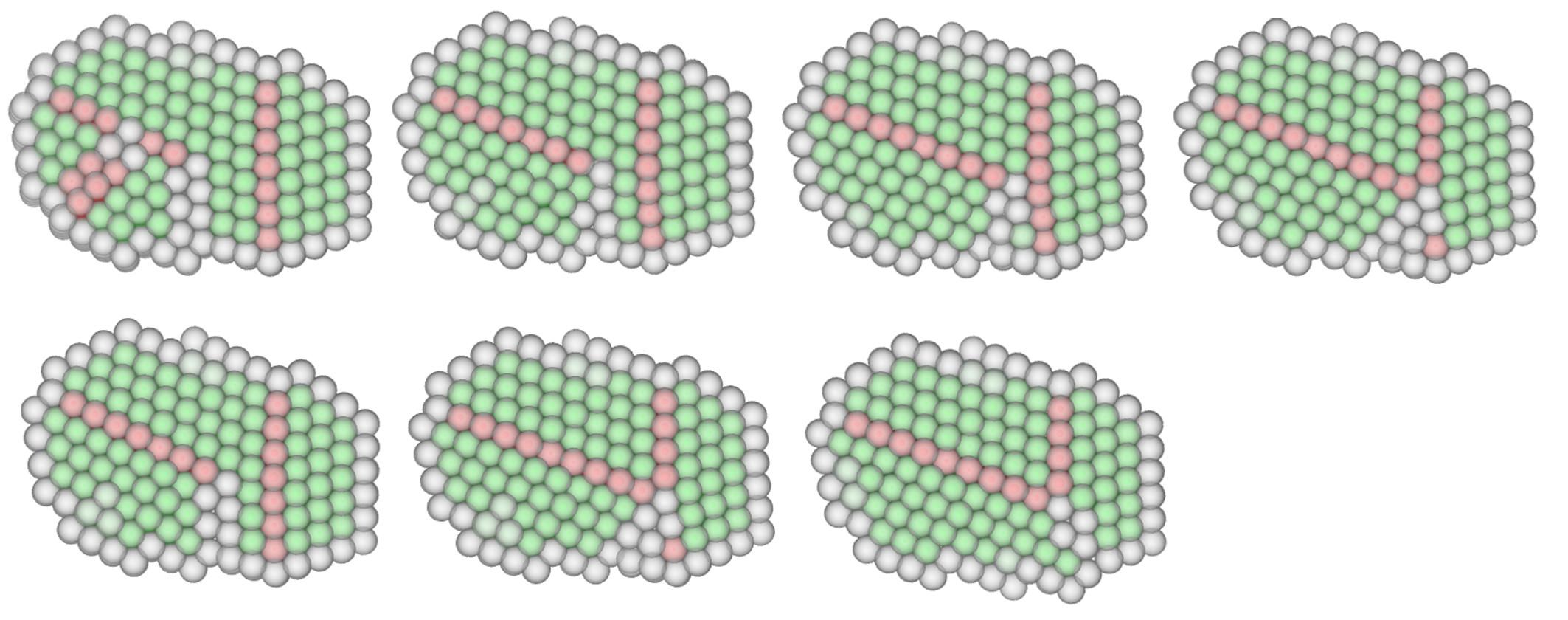}
  \caption{Structural changes during the evolution of a $\Sigma3-\Sigma11$ double junction disclination that impinges on a $\Sigma3$ GB and forms a $\Sigma3-\Sigma3-\Sigma11$ triple junction disclination.}
  \label{sup_fgr:DJ_hits_s3_to_TJ_r23}
\end{figure}

\clearpage
\section{Size effects on column shift barriers}

To analyze how the column shift barriers change with particle size, we calculated barriers using \emph{frozen-z} and NEB methods. In the \emph{frozen-z} method, several intermediate configurations are generated by applying incremental displacements to all atoms between the initial and final configurations, which differ by a column shift. For each atom, we first calculate its displacement in the final configuration with respect to the initial configuration. Each intermediate configuration is then generated by multiplying these displacements by a fractional displacement factor between 0 and 1, which is the same for all atoms. The intermediate configurations are relaxed while keeping the $z$ coordinates of all atoms fixed and allowing the $x$ and $y$ coordinates to relax. The \emph{frozen-z} method is used because, for larger columns, the full column shift can collapse into a kink mode during NEB even when restraints are applied. We emphasize that the \emph{frozen-z} method provides a qualitative comparison between the full shift and anti-parallel kink barriers, and the resulting paths are not necessarily minimum energy paths.

In Supplementary Figure~\ref{sup_fgr:barrier_flattening}a, we show the column shift barriers, defined here as the difference between the maximum and minimum energies along the path, for the full shift and anti-parallel kink modes in $\Sigma11$ and $\Sigma3$ GBs in nanoparticles and bicrystals using the \emph{frozen-z} method. The bicrystal models are finite in all three directions. Overall, the full shift barrier increases approximately linearly with column size, given by the number of atoms in the shifting column ($N_{\mathrm{col}}$). On the other hand, the anti-parallel kink barrier approaches a plateau with increasing column size.

We also tested this trend using NEB calculations. Supplementary Figure~\ref{sup_fgr:barrier_flattening}b shows the barriers for anti-parallel and parallel kink modes calculated using NEB. The anti-parallel kink barriers obtained using the \emph{frozen-z} method are also included for comparison. The plateauing of the kink barriers is clearly observed for the bicrystals, but is less obvious for the nanoparticles, presumably due to changes in surface curvature as the particle size changes, compared with the flat surfaces of the bicrystals. Nevertheless, the NEB calculations show the same overall trend as the \emph{frozen-z} calculations. For all the cases considered here, the anti-parallel kink barrier is also smaller than the parallel kink barrier.

\begin{figure}[!h]
\centering
  \includegraphics[width=0.8\textwidth]{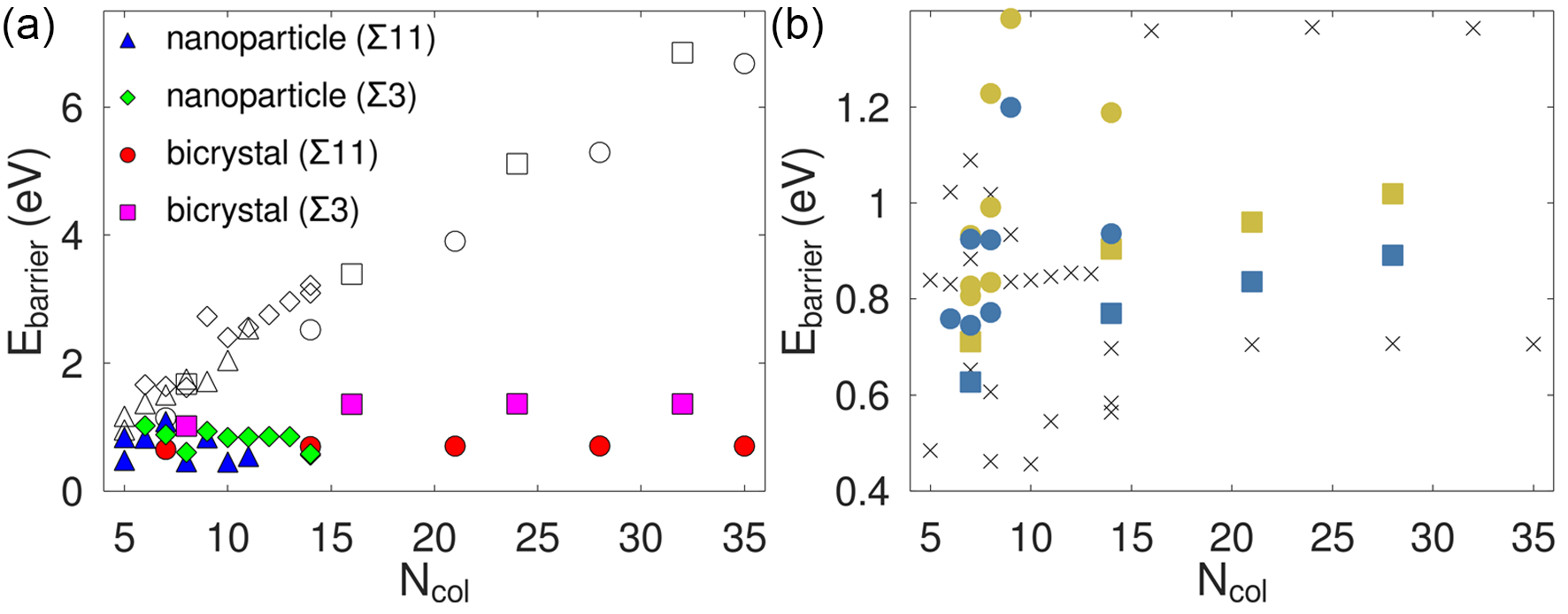}
  \caption{Energy barriers for disconnection migration through column shifts in $\Sigma11$ and $\Sigma3$ GBs in nanoparticle and bicrystal systems as a function of disconnection line length, represented by the number of atoms in the shifting column ($N_{\mathrm{col}}$), calculated using (a) \emph{frozen-z} and (b) NEB methods. The barrier is defined here as the difference between the maximum and minimum energies along the path. In (a), solid and open symbols indicate the barriers for anti-parallel kink and full shift modes, respectively. In (b), circles and squares indicate nanoparticle and bicrystal systems, respectively. Anti-parallel and parallel kink modes are shown in blue and yellow, respectively. The $\times$ markers correspond to the anti-parallel kink barriers in panel (a) calculated using the \emph{frozen-z} method.}
  \label{sup_fgr:barrier_flattening}
\end{figure}

\clearpage
\section{Effect of Ag substitution on GB migration in Cu}

Here, we present a proof of concept to show that knowledge of the C and S displacements can potentially allow targeted chemical alloying to influence GB migration. We test this idea using an ideal $\Sigma11$ GB in a Cu bicrystal, in which each grain has dimensions of 3, 4, and 14 CSL units along $\langle113\rangle$ ($x$), $\langle332\rangle$ ($y$), and $\langle110\rangle$ ($z$), respectively.

\begin{figure}[!h]
\centering
  \includegraphics[width=0.85\textwidth]{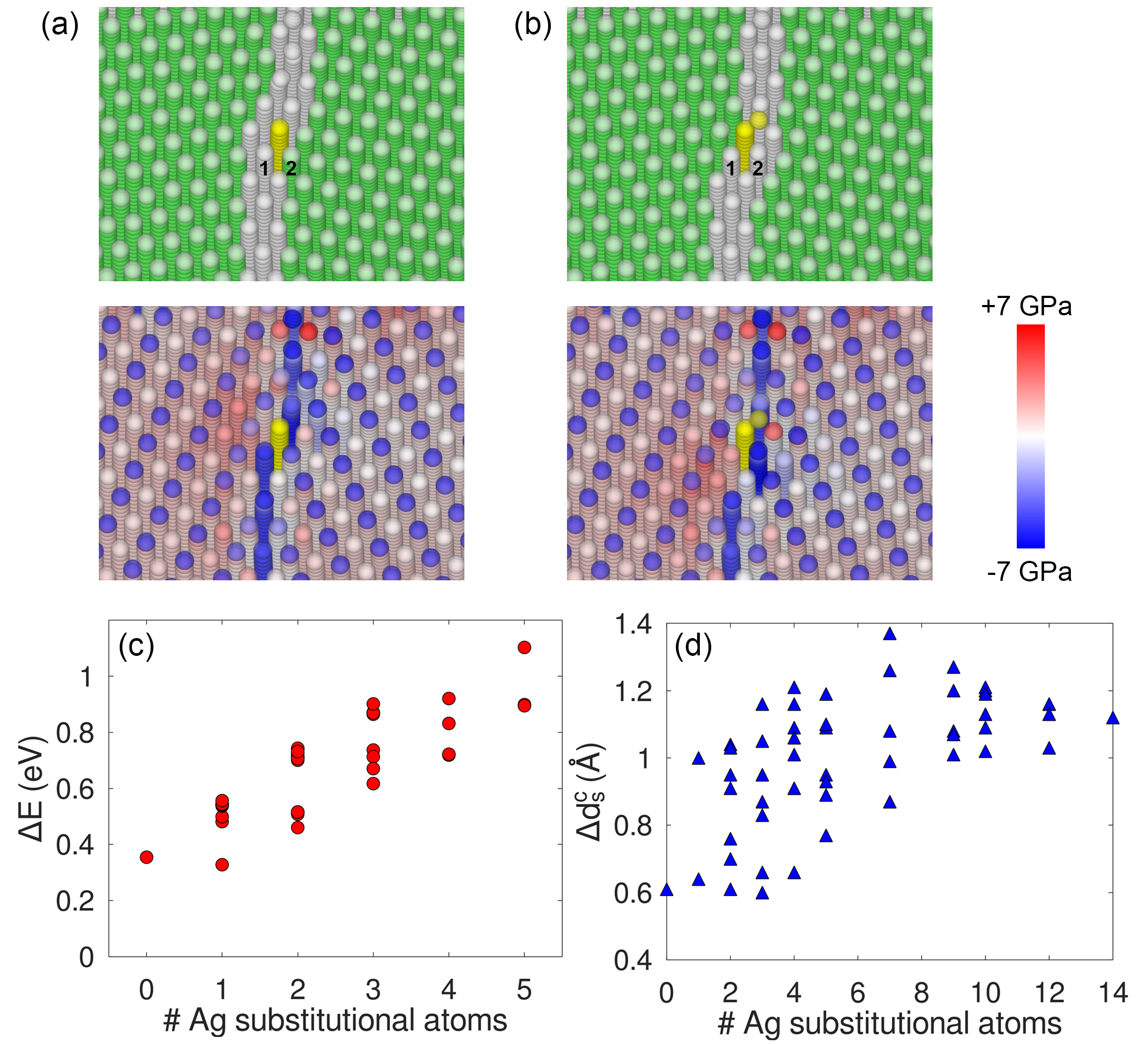}
  \caption{GB structure and corresponding atomic pressure map (a) before and (b) after the up shift of the yellow column. (c) Energy required for migration ($\Delta E$) with increasing amount of Ag in column 1. (d) Critical shear displacement (\( \mathrm{\Delta{d}}_{\mathrm{s}}^{\raisebox{0.50ex}{\scriptsize c}} \)) with increasing amount of Ag in column 1.}
  \label{sup_fgr:barrier_modification}
\end{figure}

Supplementary Figure~\ref{sup_fgr:barrier_modification}a shows the initial configuration of the Cu bicrystal, where the $\Sigma11$ GB contains a \emph{dd}, together with the corresponding atomic pressure map. After an up shift of the yellow column, the \emph{dd} migrates down by one unit (Supplementary Figure~\ref{sup_fgr:barrier_modification}b). The corresponding atomic pressure map is also shown. In both configurations, we mark columns 1 and 2, which are located just below the shifting column, where the \emph{dd} is located in the initial configuration. Before the column shift, columns 1 and 2 are below the \emph{dd}, while after the shift they are above the \emph{dd}, since the \emph{dd} has migrated down. In the initial configuration (Supplementary Figure~\ref{sup_fgr:barrier_modification}a), the atoms in column 1 have negative atomic pressure, corresponding to tensile stress, while most atoms in column 2 have atomic pressures close to zero. After the column shift, the atoms in column 2 have negative atomic pressures, while most atoms in column 1 have slightly positive atomic pressures.

We then test the effect of substituting Cu atoms in column 1 with Ag. Column 1 is a preferred location for Ag because Ag atoms are larger than Cu and can better accommodate the tensile stress at these atomic locations. Once the \emph{dd} migrates down through the column up shift, the situation reverses and the Ag atoms are now located at sites under slight compressive stress, which is less favorable for Ag. We therefore expect that Ag substitution in column 1 can inhibit the migration of the \emph{dd}.

We first calculate the energy required for migration as the amount of Ag in column 1 is increased. Column 1 consists of 14 atoms. Supplementary Figure~\ref{sup_fgr:barrier_modification}c shows that the energy required for the column shift increases on average with increasing Ag content in column 1. We also carried out a separate test in which the number of Ag atoms in column 1 was varied from 0 to 14 in steps of one, and the critical shear displacement (\( \mathrm{\Delta{d}}_{\mathrm{s}}^{\raisebox{0.50ex}{\scriptsize c}} \)) required for migration of the \emph{dd} was measured. Shear was applied using two rigid slabs at the surfaces parallel to the GB plane. The slabs were displaced in opposite directions in steps of 0.01~\AA, i.e., 0.01~\AA\ for the right slab and $-0.01$~\AA\ for the left slab, giving a total relative shear displacement of 0.02~\AA\ per step. After each displacement step, the system was relaxed while keeping the atoms in the rigid slabs fixed.

Supplementary Figure~\ref{sup_fgr:barrier_modification}d shows the variation of \( \mathrm{\Delta{d}}_{\mathrm{s}}^{\raisebox{0.50ex}{\scriptsize c}} \) with the number of Ag atoms in column 1. Unlike the migration energy, the variation is non-monotonic. The critical shear displacement is higher with 14 Ag atoms than with no Ag in column 1. In between, \( \mathrm{\Delta{d}}_{\mathrm{s}}^{\raisebox{0.50ex}{\scriptsize c}} \) increases up to around 8 Ag atoms and then decreases towards the value at 14 Ag atoms.

Overall, this example shows that knowledge of the GB migration mechanism, in this case migration of a \emph{dd} through a column shift, together with the local atomic pressure distribution can be used to target changes in GB migration through local chemistry. For example, Supplementary Figure~\ref{sup_fgr:barrier_modification}d shows that the largest critical shear displacement occurs around 8 Ag atoms rather than when the entire column is substituted by Ag.

%\newpage
%\begin{figure}[!t]
%\centering
%  \includegraphics[width=1.0\textwidth]{figures/png_ns3_sup_fig_4J_detach_to_s3_r30.png}
%  \caption{zDiff.}
%  \label{sup_fgr:4J_to_s3_r30}
%\end{figure}

\end{document}